\documentclass[useAMS,usenatbib,usegraphicx]{mn2e}
\usepackage{amsmath}
\usepackage{caption}
\usepackage{amssymb}
\usepackage{color}
\usepackage{xspace}
\usepackage{soul}
\usepackage{verbatim}

\pubyear{2010}

\newcommand{\be}{\begin{equation}}
\newcommand{\ee}{\end{equation}}
\newcommand{\ba}{\begin{eqnarray}}
\newcommand{\ea}{\end{eqnarray}}

\newcommand{\al}{\alpha}

\newcommand{\OM}{\Omega_{\rm M}}
\newcommand{\OL}{\Omega_\Lambda}

\begin{document}
\title[Supervoids in WISE-2MASS and Cold Spots in the CMB]
{Supervoids in the WISE-2MASS catalogue imprinting Cold Spots in the Cosmic Microwave Background}

\author[F. Finelli, J. Garc\'ia-Bellido, A. Kov\'acs, F. Paci, I. Szapudi]
{F.~Finelli $^{1,2}$, J. Garc\'ia-Bellido $^{3}$\thanks{E-mail: juan.garciabellido@uam.es}, A. Kov\'acs $^{4}$, 
F. Paci $^{5}$, I. Szapudi $^{6}$ 
\\
$^1$ INAF-IASF Bologna, Istituto di Astrofisica Spaziale e Fisica Cosmica di Bologna \\
Istituto Nazionale di Astrofisica, via Gobetti 101, I-40129 Bologna, Italy \\
$^2$ INFN, Sezione di Bologna, Via Irnerio 46, I-40126 Bologna, Italy\\
$^3$ Instituto de F\'isica Te\'orica IFT-UAM/CSIC, Universidad Aut\'onoma de Madrid, Cantoblanco 28049 Madrid, Spain\\
$^4$ Institut de F\'isica d'Altes Energies, Universitat Aut\'onoma de Barcelona, E-08193 Bellaterra (Barcelona), Spain\\
$^5$ SISSA, Astrophysics Sector, Via Bonomea 265, 34136, Trieste, Italy\\
$^6$ Institute for Astronomy, University of Hawaii, 2680 Woodlawn Drive, Honolulu, HI, 96822, USA}

\label{firstpage}

\maketitle

\begin{abstract}
The Cold Spot (CS) is a clear feature in the Cosmic Microwave 
Background (CMB); it could be of primordial origin, or caused 
by a intervening structure along the line of sight. We identified a large projected 
underdensity in the recently constructed WISE-2MASS all-sky 
infrared galaxy catalogue aligned with the Cold Spot direction 
at $(l,b)\approx(209^\circ,-57^\circ)$. It has an angular size 
of tens of degrees, and shows a $\sim20\%$ galaxy underdensity 
in the center. Moreover, we find another large underdensity in 
the projected WISE-2MASS galaxy map at $(l,b)\approx(101^\circ,46^\circ)$ 
(hereafter Draco Supervoid), also aligned with a CMB decrement, although 
less significant than that of the CS direction. Motivated by these findings, 
we develop spherically symmetric Lemaitre-Tolman-Bondi (LTB) compensated void models to 
explain the observed CMB decrements with these two underdensities, or ``supervoids''. 
Within our perturbative treatment of the LTB voids,  we find that the Integrated 
Sachs-Wolfe and Riess-Sciama effects due to the Draco Supervoid can account for 
the CMB decrement observed in the same direction.  On the contrary, the extremely 
deep CMB decrement in the CS direction is more difficult to explain by the presence 
of the CS supervoid only. Nevertheless, the probability of a random alignment between 
the CS and the corresponding supervoid is disfavored, and thus its contribution as a 
secondary anisotropy cannot be neglected. We comment on how the approximations used in 
this paper, in particular the assumption of spherical symmetry, could change quantitatively 
our conclusions and might provide a better explanation for the CMB CS.
\end{abstract}


\section{Introduction}

The temperature anisotropies of the Cosmic Microwave Background (CMB) provide 
the earliest image of the primordial density fluctuations generated during inflation. 
They evolve into the present distribution of Dark Matter traced by
galaxies, possibly in a biased fashion. 
The cosmological information encoded in the spatial distribution of galaxies 
is revealed by several present and future programs mapping the Universe in 
wide area surveys, such as Pan-STARRS~\citep{kaiser2010},  DES\footnote{http://www.darkenergysurvey.org/},
DESI~\citep{DESI}, LSST~\citep{LSST} and Euclid~\citep{euclid_redbook,euclid}.

While a homogeneous and flat Universe with a cosmological constant and dark matter - the $\Lambda$CDM model - 
agrees with most of the observations,  several ``anomalies'' remain puzzling on large angular scales in the CMB maps. Among these, a cold area in the direction 
$(l,b)=(209^\circ, -57^\circ)$, initially found as of approximatively $5^\circ$ radius in the Wilkinson Microwave Anisotropy Probe 
(WMAP) temperature map by \cite{Vielva2004}, has been characterized in the {\sc Planck} 2013 
temperature maps with great accuracy \citep{PlanckIandS}. The occurrence of a decrement in the 
CMB temperature pattern with similar size in Gaussian simulations has been 
evaluated to be below 1\%  \citep{Cruzetal2005}. 
Although the statistical significance of the CS as an anomaly in the CMB pattern 
within the $\Lambda$CDM cosmology can be debated, it is important to pursue physical explanations beyond 
assuming it is astatistical fluke.

\begin{figure*}
\includegraphics[width=\textwidth]{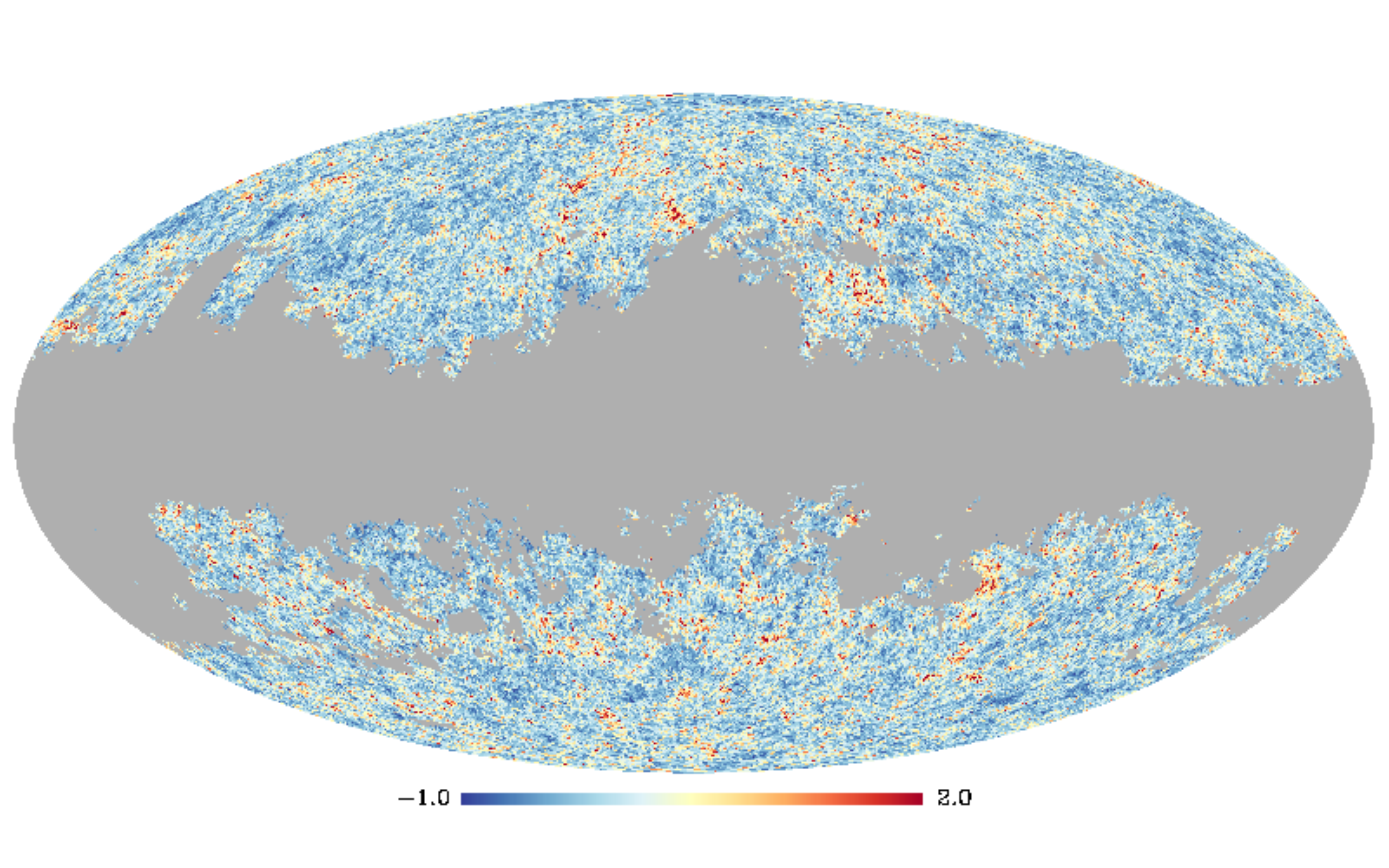}
\caption{WISE-2MASS galaxy density map with the corresponding mask.  See text for details.}
\label{w2m_full}
\end{figure*}

The CS in the CMB pattern could have been originated by
a primordial fluctuation on the last scattering surface or by an intervening
phenomenon along the line of sight. Contamination from our galaxy or 
by the Sunyaev-Zeldovich effect from a cluster are quite unlikely \citep{cruztucci}.
If it were a primordial feature on the last scattering surface, the CS could be a signature 
of a non-perturbative effect during inflation \citep{GBH2008,afshordi}, or it might open a new 
exciting window onto the early Universe if caused 
by a cosmic texture generated during a phase transition at $10^{16}$ GeV \citep{cruzscience}.
Alternatively, the CS could be imprinted by an intervening supervoid along the line of 
sight \citep{Inoue2006,Inoue2007,Inoueetal2010,Inoue2012}.
While voids fill approximately 30 \% of the Universe at $z<1$ \citep{colberg}
a quantitative explanation seems to require a supervoid with radius $\gtrsim 200 \, {\rm Mpc}^{-1}$,
a quite rare structure in a $\Lambda$CDM cosmology. Future data 
from CMB lensing~\citep{DS} and from 21-cm observations~\citep{KK} may help to further
discriminate between the texture and the void hypotheses.

The supervoid hypothesis can be tested in galaxy surveys, and several investigations have 
been already carried out. An underdensity in the direction of the CMB cold spot has been claimed in 
NVSS \citep{Rudnicketal2007}, but its statistical significance has been debated \citep{SmithHuterer2010}.
\cite{Granettetal2010} imaged the cold spot region by the Canada-France-Hawaii Telescope (CFHT) ruling 
out  the existence of a 100 Mpc supervoid with underdensity 
$\delta \simeq -0.3$ at $0.5 < z < 0.9$. \cite{Bremeretal2010} 
reached similar conclusions from a redshift survey using the 
VIMOS spectrograph on the VLT. In the relatively shallow 2MASS 
galaxy catalogue, \cite{francis2010} found an under-density in the 
galaxy field in the Cold Spot region. The structure they identified induces a $\Delta T=-7 \mu$K 
depression in the CMB temperatures in the $\Lambda$CDM cosmology, that is not a satisfactory explanation for the CS anomaly.

In this paper and in \cite{SzapudiEtal2014} a supervoid is indeed found
in the direction of the CMB CS at a redshift compatible with previous constraints
\citep{Bremeretal2010,Granettetal2010}.
In this paper we characterize the underdensity with the use of the
recent galaxy catalogue WISE-2MASS \citep{KovacsandSzapudi2013}
produced by joining the Wide-Field Infrared Survey Explorer (WISE, \cite{Wright2010})
with the 2-Micron All-Sky Survey (2MASS, \cite{Skrutskie2006}), whose median
redshift is $z \sim 0.14$. (See \cite{Kovacs2013} for the previous generation of catalogue based on WISE alone.)
In \cite{SzapudiEtal2014} a joint WISE-2MASS-Pan-STARRS1 catalog constructed within
a $50^\circ\times50^\circ$ area centred on the CMB CS with photometric redshifts is used to best characterize
the supervoid and determine also its deepest region at $z \sim 0.15 - 0.25$.

Motivated by this detection of a supervoid in the direction of the CMB CS
we use a Lemaitre-Tolman-Bondi model to jointly explain the underdensity in the WISE-2MASS galaxy survey 
and the decrement in the CMB temperature maps. Since a correspondence between supervoids and cold regions 
in the CMB temperature pattern could not be unique to the CS direction, we search for other large underdensities 
in the WISE-2MASS catalogue, extending \cite{SzapudiEtal2014} who surveyed only the CS area. 
We find {\em another} underdensity located in the Northern Galactic hemisphere in the direction 
at $(l\approx101^\circ \,, b\approx46^\circ)$ with an angular size similar to the one found in 
the direction of the CMB CS. While this projected underdensity is even deeper than the one found 
in the CS region, the shallow 2MASS galaxy density maps of $z_{mean}=0.07$ by \cite{Rassat2013} 
show a large underdensity in a similar location, therefore it may be closer to us and smaller 
in physical size. Hereafter we refer to this underdensity as Draco Supervoid.
Since this underdensity is aligned with a moderate CMB temperature decrement as well, 
we apply our LTB model to jointly explain the galaxy and CMB pattern in this direction.

The paper is organized as follows. In section 2 we briefly describe the {\sc Planck} nominal mission and WMAP 9-year temperature data which we use for our analysis, and also describe the WISE-2MASS data and the two largest underdensities found.
In section 3, we describe our basic LTB model for voids and the corresponding predictions for CMB temperature 
anisotropies, and apply our basic LTB model to the underdensities found in the WISE-2MASS catalogue. We draw our conclusions in section 4.

\section{Data sets}

\subsection{CMB temperature maps}

As CMB temperature maps we consider WMAP 9-year and {\sc Planck} data. 
To minimize astrophysical contamination
we use CMB foreground cleaned maps. For WMAP, we use the 9th year Internal Linear Combination
(ILC) map at the HEALPIX \citep{gorski} resolution $N_{\rm side}=1024$
publicly provided at http://lambda.gsfc.nasa.gov/.
For {\sc Planck} data different CMB foreground cleaned maps are provided \cite{PlanckCompSep},
and we choose the Spectral Matching Independent Component Analysis (SMICA) \citep{Cardosoetal2008}
map at the HEALPIX \citep{gorski} resolution $N_{\rm side}=2048$,
publicly provided at http://pla.esac.esa.int/pla/aio/planckProducts.html.

\subsection{WISE-2MASS galaxies}

\cite{KovacsandSzapudi2013} combined
photometric information of the WISE and 2MASS
infrared all-sky surveys to produce a clean galaxy sample for large-scale structure
research. They apply Support Vector Machines (SVM)
to classify objects using the multicolor WISE-2MASS database. They
calibrate their star-galaxy separator algorithm using Sloan Digital Sky
Survey \citep[SDSS,][]{SDSS} classification, and use the Galaxy and Mass
Assembly \citep[GAMA,][]{gama} spectroscopic survey for determining the
redshift distribution of the WISE-2MASS galaxy sample.

Furthermore, \cite{KovacsandSzapudi2013} pointed out that $W1_{\rm WISE} - 
J_{\rm 2MASS} \leq -1.7$ with a flux limit of $W1_{\rm WISE}$ $\leq$ 15.2 mag is a 
simple and effective star-galaxy separator, capable of
producing results comparable to the multi-dimensional SVM classification. As a further refinement,
another flux limit of $J_{\rm 2MASS}$  $\leq$ 16.5 mag is applied to have a fairly uniform all-sky
dataset that is deeper than 2MASS XSC \citep{JarrettEtAl2000} and cleaner the WISE.
The final catalogue has an estimated $\sim 2\%$ stellar contamination among
2.4 million galaxies with $z_{med}\approx 0.14$.

We construct a mask to exclude potentially contaminated regions near the
Galactic plane using the dust emission map of \cite{schlegel_DUST}. We mask
out all pixels with $E(B-V) \geq 0.1$, and regions at galactic latitudes
$|b| < 20^{\circ}$, leaving 21,200 ${\rm deg}^{2}$ for our purposes. 
The resulting galaxy density map with the corresponding mask is shown in Fig.\ref{w2m_full}.

\subsubsection{Cold Spot region in WISE-2MASS}

\begin{figure}
  \centering
\includegraphics[width=0.9\columnwidth,angle=0]{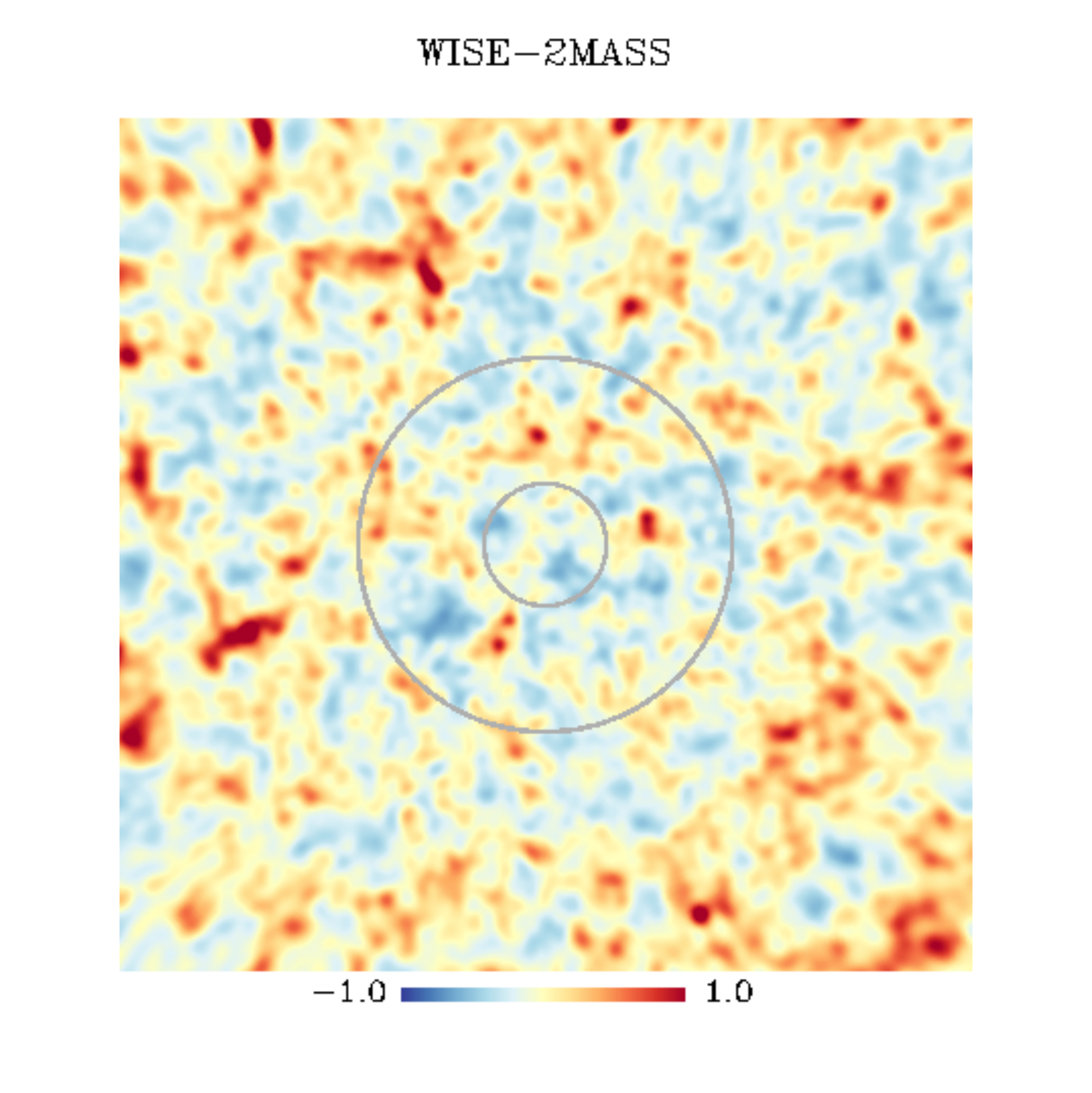}
\includegraphics[width=0.9\columnwidth,angle=0]{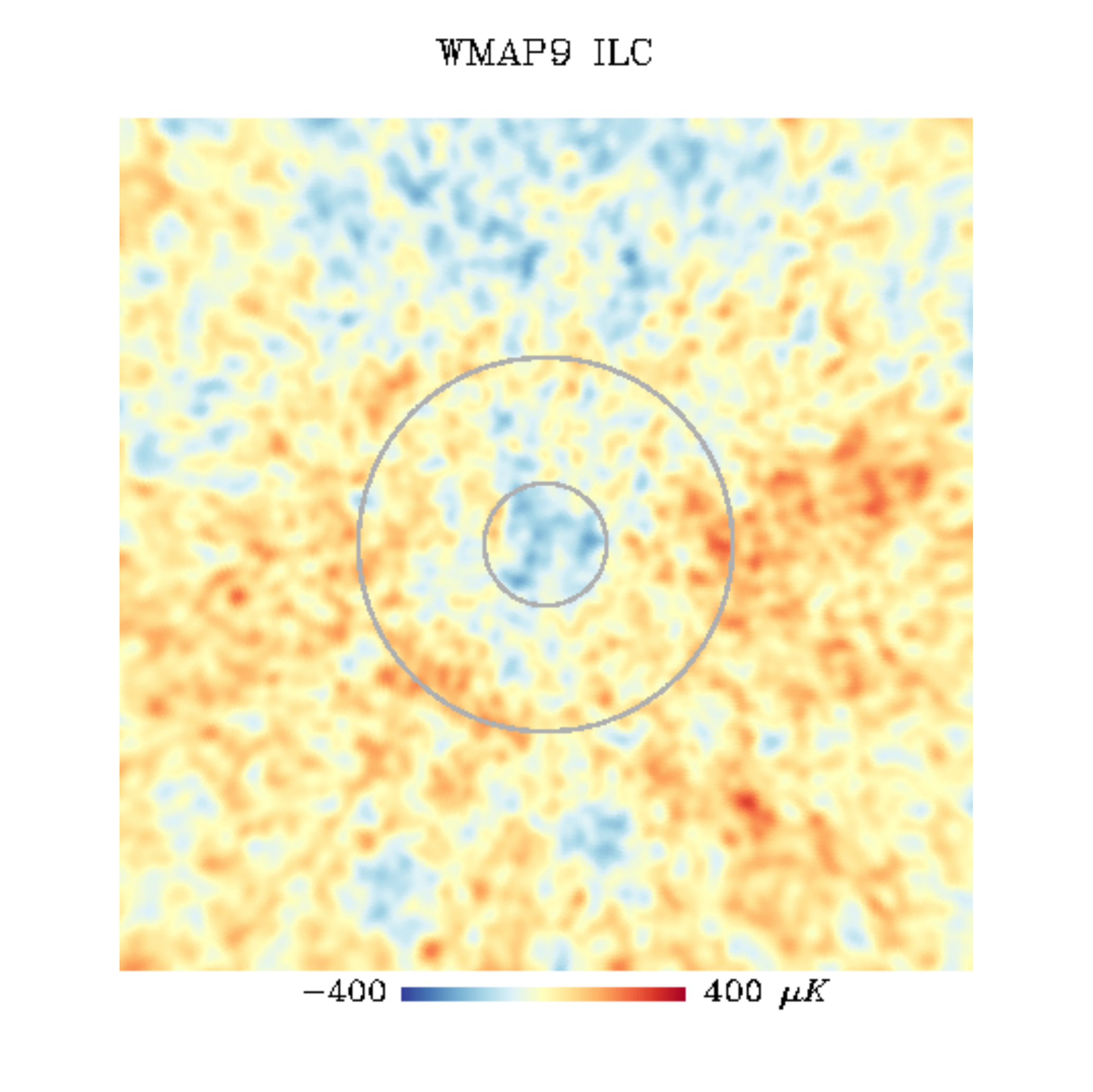}
\includegraphics[width=0.9\columnwidth,angle=0]{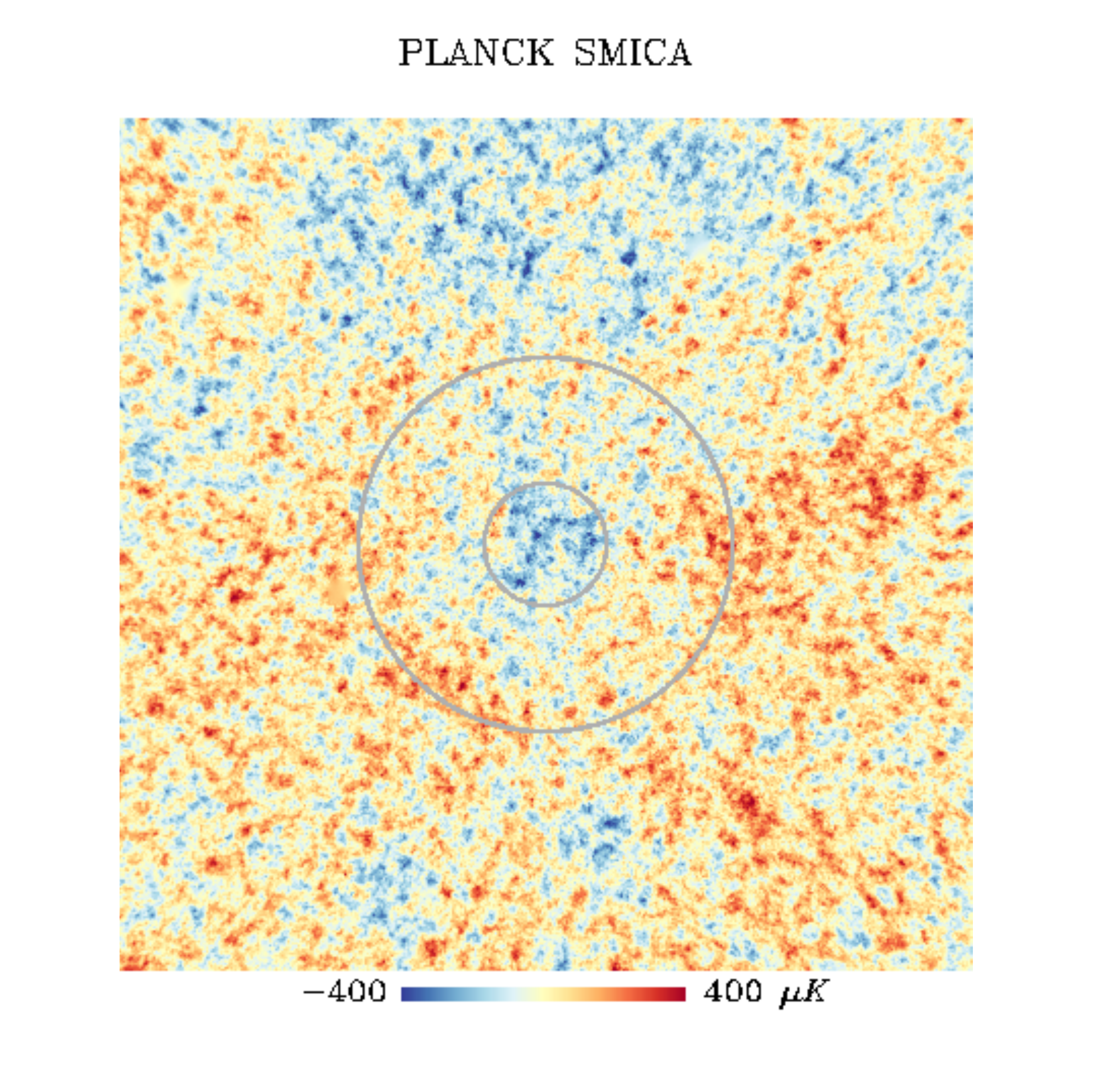}
\caption{The WISE-2MASS (top), WMAP9 (middle), and {\sc Planck} SMICA (bottom) field in the direction of the Cold Spot. Circles correspond to $5^{\circ}$ and $15^{\circ}$ radii.}
\label{CS_maps}
\end{figure}

We first study the CMB Cold Spot region on the sky as traced by WISE-2MASS galaxies in the $\sim 30^\prime$ resolution 2D projection - {\tt HEALPix} \citep{gorski} 
 $N_{\rm side}=128$ -, 
therefore perform a complementary analysis to that of \cite{SzapudiEtal2014} who measured galaxy densities in the line-of-sight in their WISE-2MASS-PS1 photo-z data.
Top panel of Fig. \ref{CS_maps} shows a $60^\circ\times 60^\circ$ size patch of the galaxy catalogue, centered on the CS direction.
At the center of the CS, there is an approximately 20\% underdensity in galaxy 
counts, extending for a radius over $20^\circ$. We observed that the deepest part of this underdensity is slightly off-center with respect to the center of the Cold Spot estimated based on CMB data, but we only use this nominal center in our analysis to minimize posteriori choices.

The image in the WISE-2MASS galaxy catalogue is compared to the same angular patch of  
the WMAP 9-year Internal Linear Combination (ILC) map \citep{bennet2012} and of the {\sc Planck}~SMICA map \citep{PlanckCompSep} in Fig.~\ref{CS_maps}. An average 70 $\mu$K temperature decrement of approximatively 10 degree 
in size is clearly visible, corresponding to the CMB Cold Spot \citep{Cruzetal2005}. 

\subsubsection{Further large voids in WISE-2MASS?}

\begin{figure}
\includegraphics[width=\columnwidth]{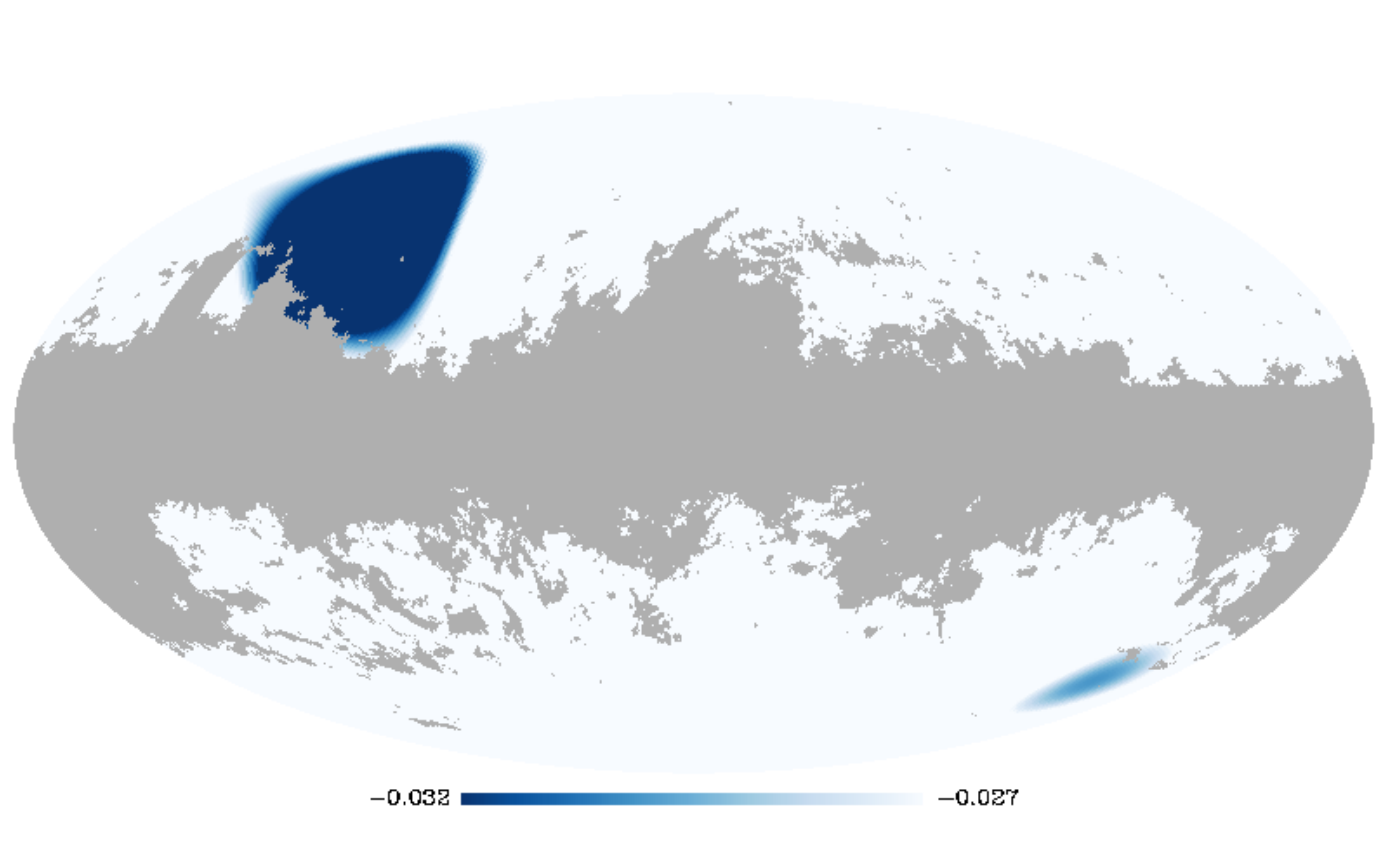}
\caption{Applying a $\sigma=20^{\circ}$ Gaussian smoothing 
to the WISE-2MASS galaxy map one finds two large underdensities 
on the full observable sky. One of them is the famous Cold Spot region,
the other one is the Draco supervoid. 
} 
\label{fullsky}
\end{figure}

\begin{figure}
\centering
\includegraphics[width=0.9\columnwidth,angle=0]{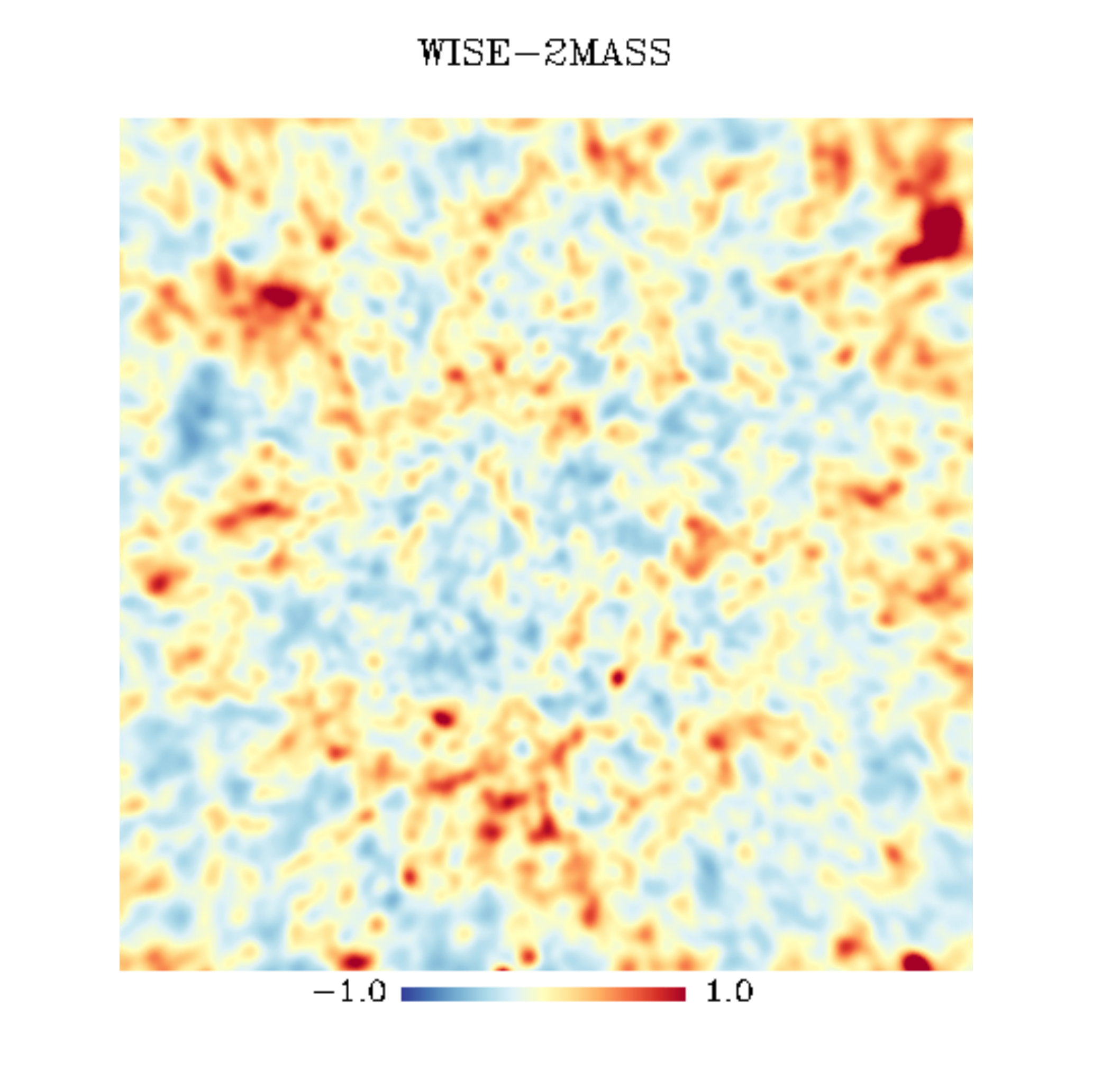}
\includegraphics[width=0.9\columnwidth,angle=0]{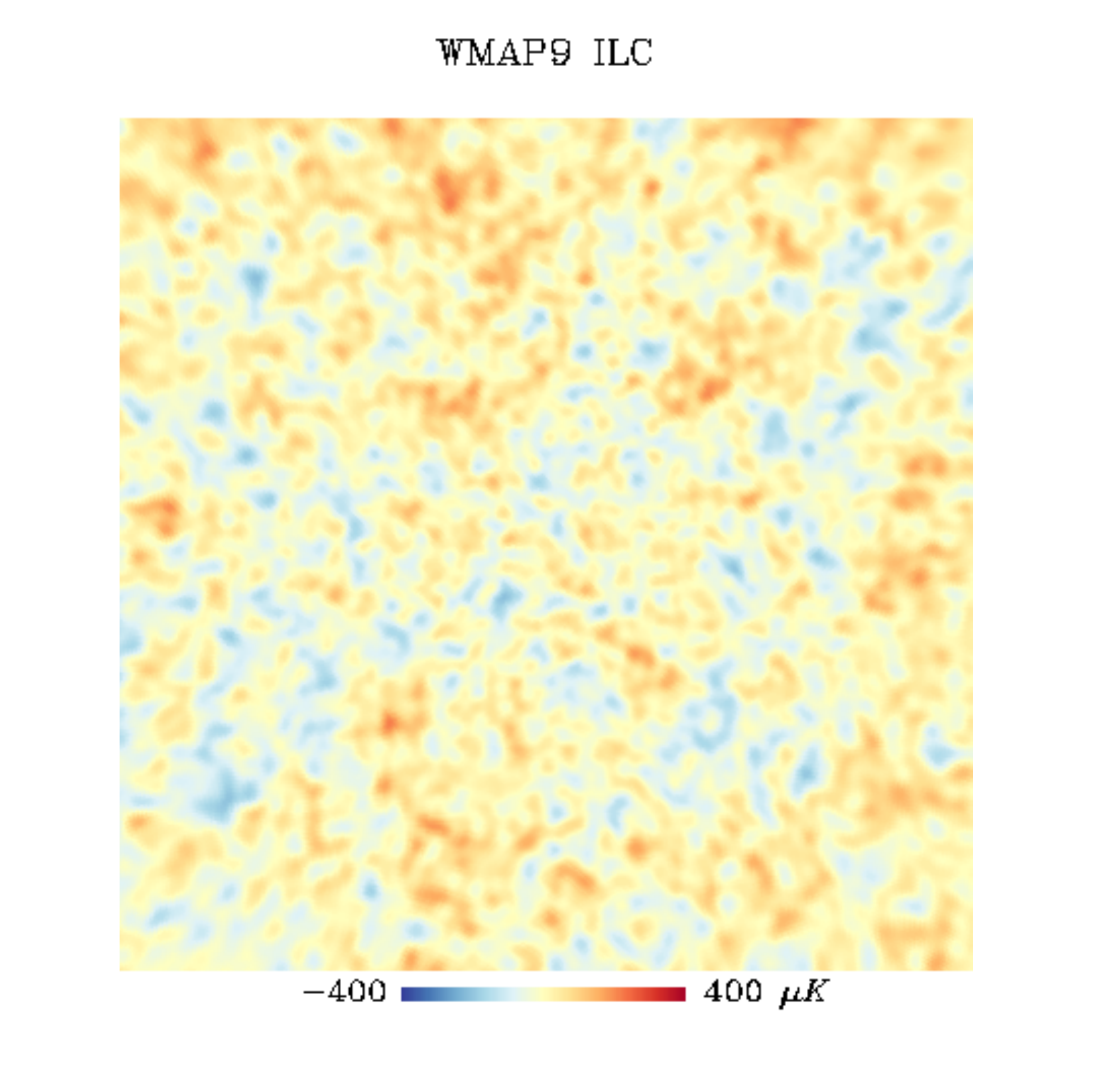}
\includegraphics[width=0.9\columnwidth,angle=0]{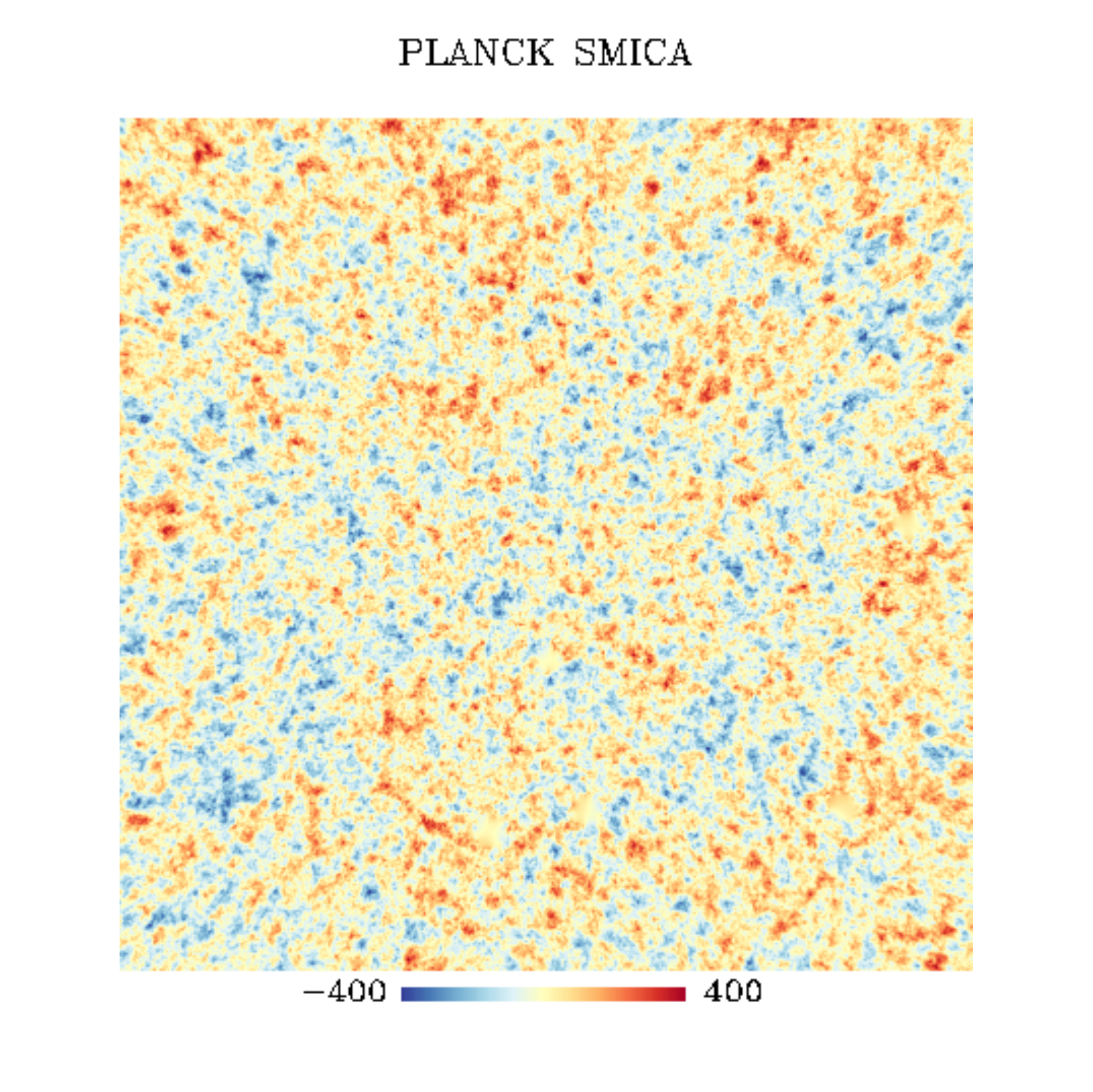} \\
\caption{The WISE-2MASS (top), WMAP9 (middle), and {\sc Planck} SMICA (bottom) field in the direction of the Draco void.}
\label{Draco_maps}
\end{figure}

We proceed with identifying underdensities in WISE-2MASS that are 
comparable to that of the CMB Cold Spot region. The galaxy density field was 
smoothed with a $\sigma=20^{\circ}$ Gaussian kernel to filter out all underdensities smaller than the 
angular radius of the supervoid found at the CS. All pixels as ``cold" as the CS supervoid region have 
been selected. Our findings are summarized in Fig.~\ref{fullsky}. There is another large underdensity 
in the WISE-2MASS catalog that has similar angular size to the CS. Moreover, this underdensity 
is somewhat deeper in its center, as shown in Fig. \ref{v2_prof}. We deem the 
center of the void as the coldest pixel's location at $\ell \approx 101^{\circ}$, $b \approx 46^{\circ}$,
and for simplicity we call it the Draco Supervoid to distinguish it form the CS void. 
The $60^\circ\times 60^\circ$ size patch of the galaxy and the CMB field corresponding to the Draco Supervoid is shown in Fig. \ref{Draco_maps}.


\begin{figure*}
\centering
\includegraphics[width=1.\columnwidth,angle=0]{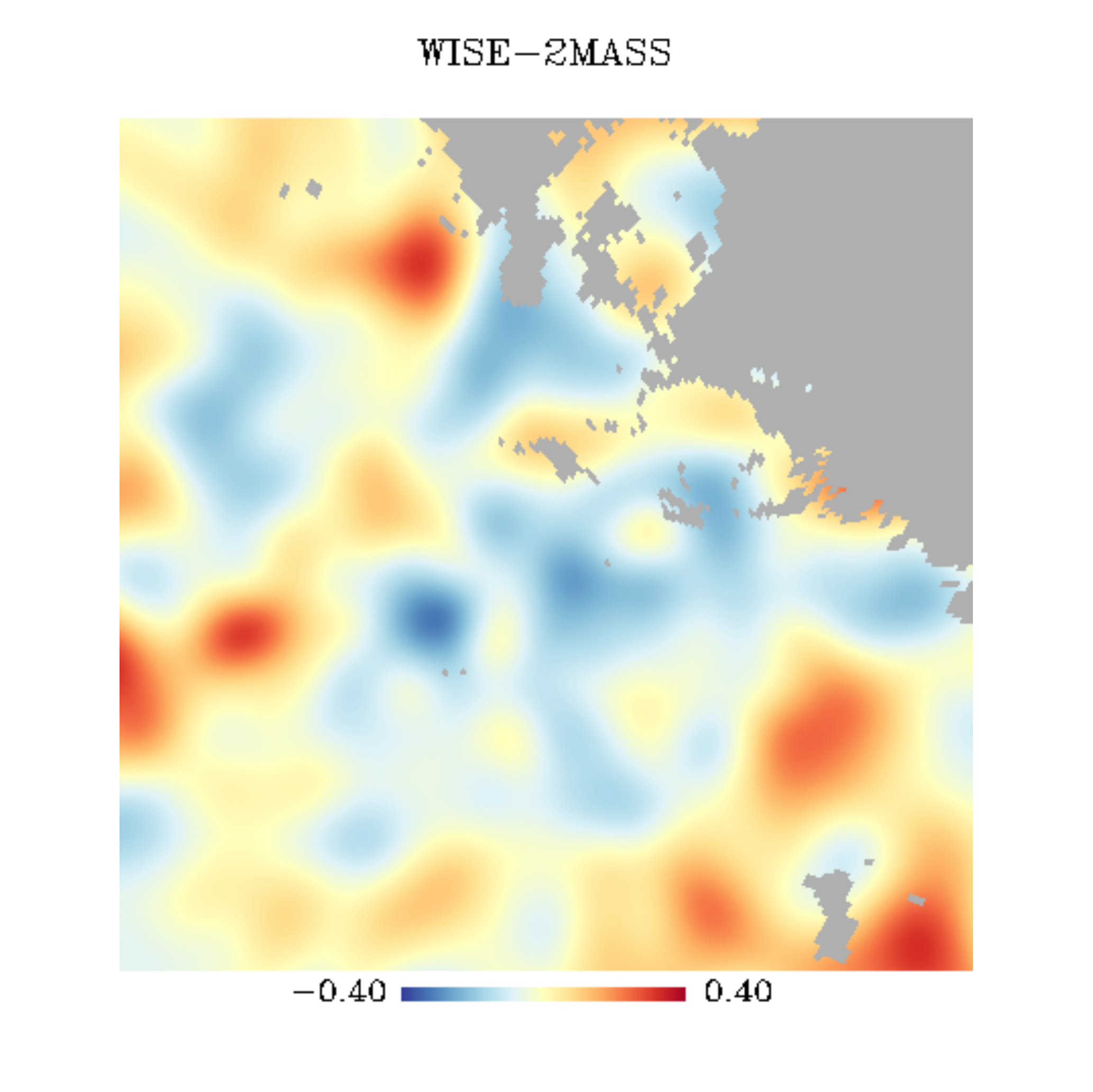} 
\includegraphics[width=1.\columnwidth ,angle=0]{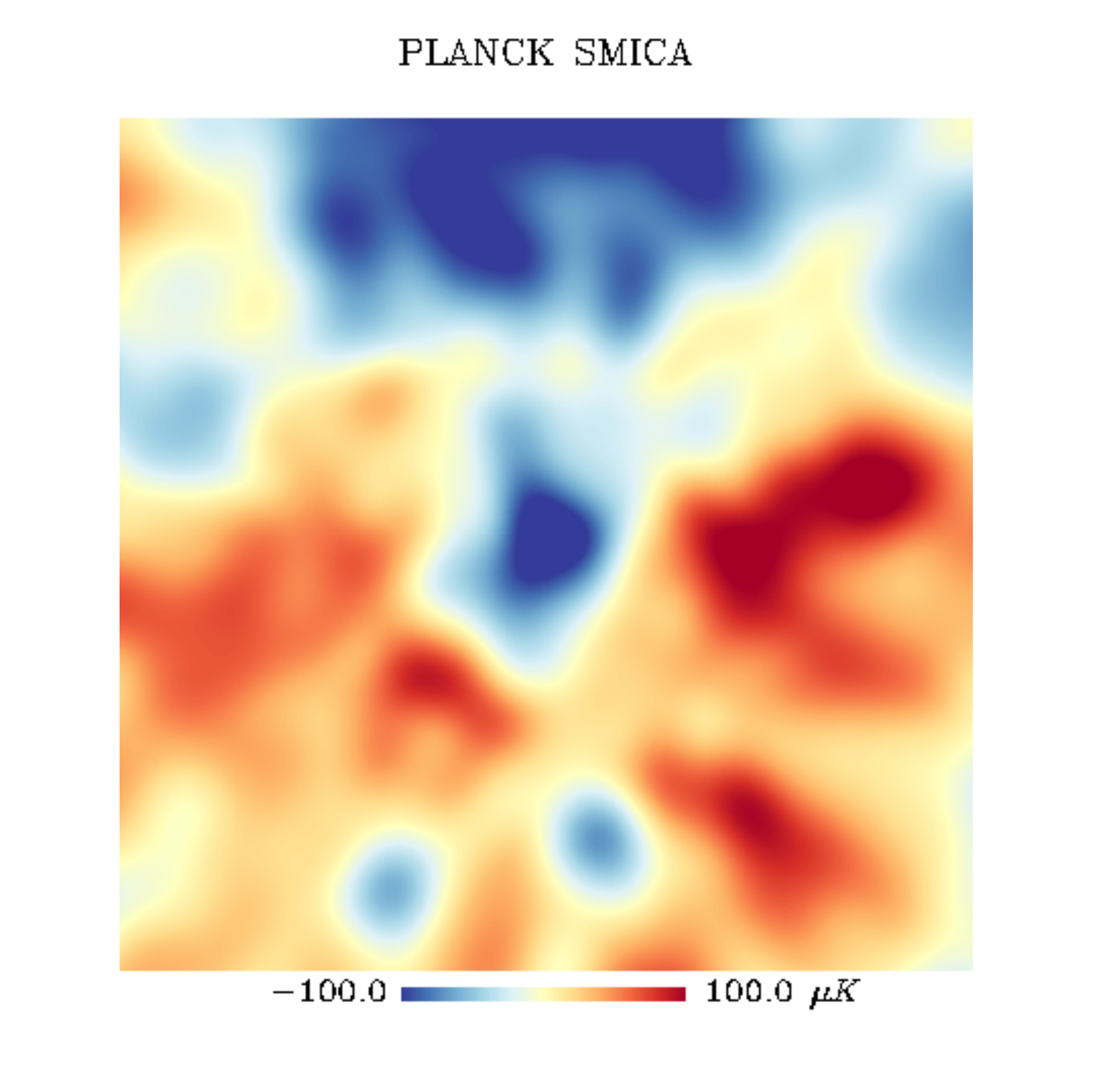} \\
\includegraphics[width=1.\columnwidth,angle=0]{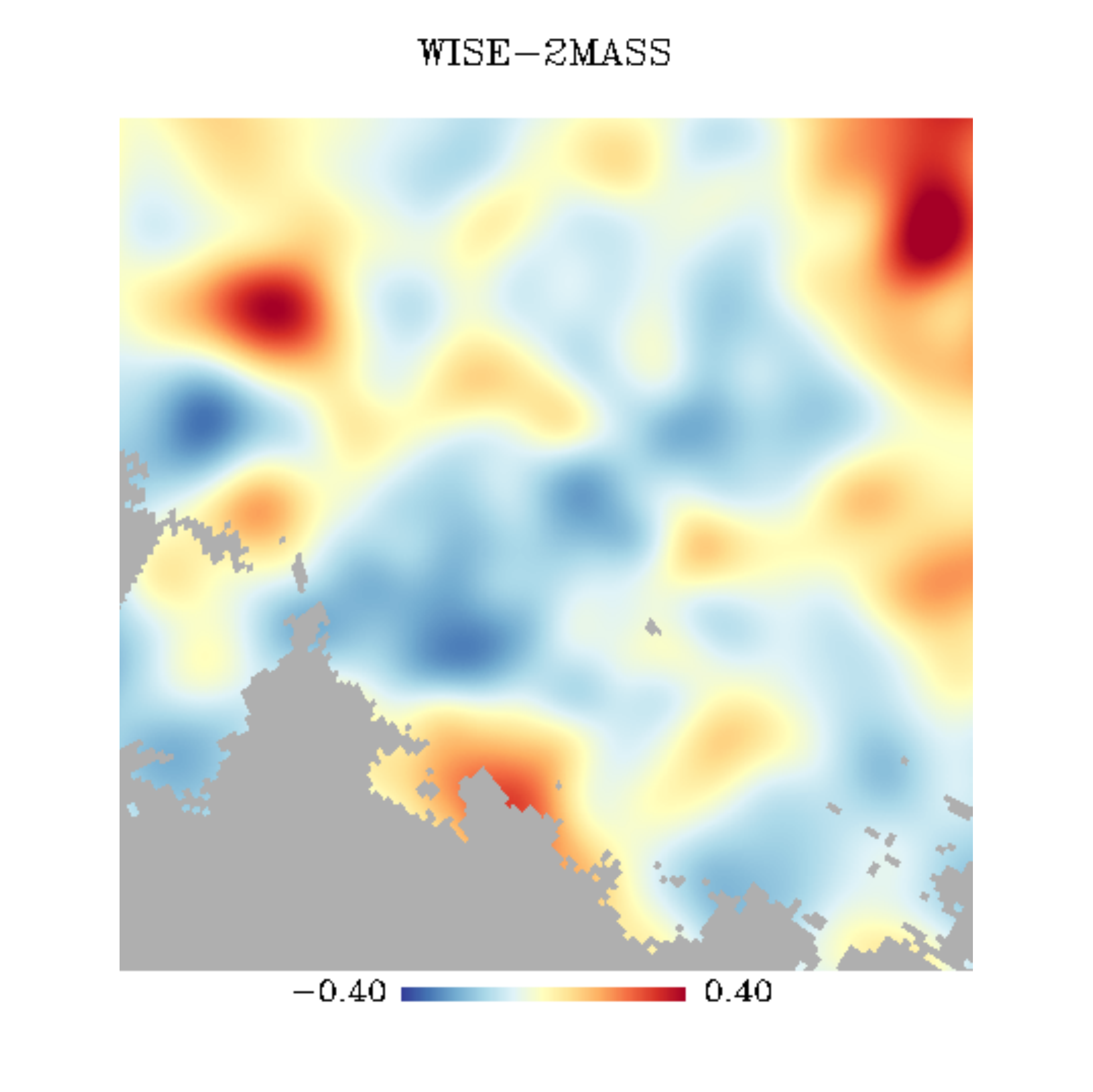}
\includegraphics[width=1.\columnwidth,angle=0]{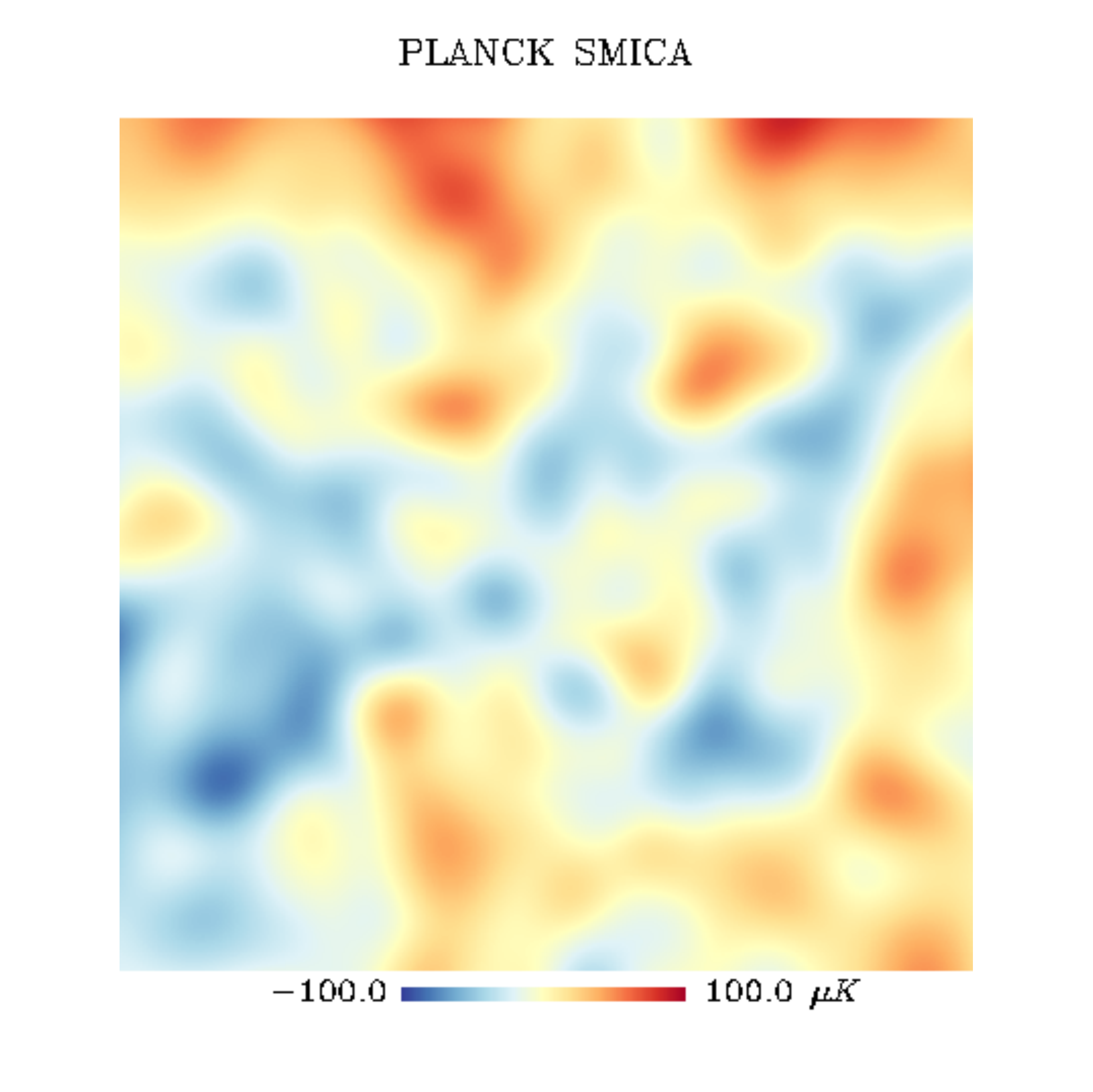}
\caption{Top panels: The $5^{\circ}$ Gaussian smoothed WISE-2MASS (left), and {\sc Planck} SMICA (right) field in the direction of the Cold Spot.  Bottom panels: same data in the direction of the Draco supervoid. The grey area in the WISE-2MASS panels shows the Galactic region we mask out for our analysis.}
\label{smoothed_maps}
\end{figure*}

A 5-degree Gaussian smoothed version of the CS and the Draco void field is shown in Fig. \ref{smoothed_maps}.

\section{The Basic $\Lambda$LTB  Void Model}

We model an underdensity in WISE-2MASS as a compensated void profile within a 
$\Lambda$LTB model, \citep{GBH2008}, 
\be
ds^2 = - dt^2 + \frac{A'(r,t)^2}{1-k(r)}\,dr^2 + A(r,t)^2\,d\Omega^2\,,
\ee
characterized by a spatial curvature profile given by 
\be
k(r)=k_0\,r^2 \exp\Big[\!-\frac{r^2}{r_0^2}\Big] \,.
\label{profilekr_basic}
\ee
See also \cite{Romano:2013kua} for a recent use of a Gaussian profile for LTB models in another context.
We approximate the novel $\Lambda$LTB model 
introduced above as a linear perturbation $\Phi$ in the synchronous gauge
in a Friedman-Robertson-Walker (FRW) metric with $\OM+\OL=1$:
\ba
\Phi(\tau,r) &\!=\!& \Phi_0(r)\cdot \ {}_2\!F_1\Big[1,\,\frac{1}{3},\,\frac{11}{6},-\frac{\OL}{\OM}a^3\Big] \,,
\label{Phi_basic} \\
&\! \equiv \!& \Phi_0 \, \exp\Big[\!-\frac{r^2}{r_0^2}\Big] \cdot F_1(u) \,
\label{profilePhi_basic}
\ea
where $u=\frac{\OL}{\OM}a^3$ and $\tau$ is the FRW conformal time: 
\be
\tau(a) = \frac{2\sqrt a}{H_0\sqrt\OM}\cdot   \label{tau}
{}_2\!F_1\Big[\frac{1}{2},\,\frac{1}{6},\,\frac{7}{6},-\frac{\OL}{\OM}a^3\Big]
\equiv \frac{2\sqrt a}{H_0\sqrt{\OM}}\cdot F_2(u)\,.
\ee

The scalar potential (\ref{profilePhi_basic}) gives rise to a 3D density profile for the void, via Poisson equation 
$\nabla^2\Phi = \frac{3}{2}H_0^2\OM\,\delta/a$,
\be
\delta(\tau,r) =  - \delta_0\,g(\tau) \Big( 1 - \frac{2}{3}\,\frac{r^2}{r_0^2} \Big) \, \exp\Big[\!-\frac{r^2}{r_0^2}\Big]\,,
\label{profile3D}
\ee
characterized by two parameters, the comoving width of the void, $r_0$ \footnote{In a first version 
of this manuscript we followed \cite{MasinaNotari2009} and considered a large difference 
between the FRW and the $\Lambda$LTB radii. This difference was due to an incorrect matching 
of the $\Lambda$LTB void to an Einstein-de Sitter model with $\OM=1$, which was performed 
by \cite{BiswasNotari2007} for a different physical problem.
The FRW and $\Lambda$CDM radii are now approximatively the same as in \cite{Zibin,Nadathuretal}.}, 
and its depth today, $\delta_0$.
The following relation between $\Phi_0$, $k_0$ and $r_0$ holds:
\be\label{phi0}
\Phi_0 = \frac{-3k_0 r_0^2}{40} = \frac{\OM}{4}\frac{H_0^2\,r_0^2\,\delta_0}{F_1(-\OL/\OM)}\,.
\ee
We write the density contrast growth factor $g(a)$ as:
$$g(a) \equiv \frac{\delta(a)}{\delta(1)} = \frac{a\cdot F_1(u)}{F_1(-\OL/\OM)}\,.$$
It is easy to check that the 3D density profile (\ref{profile3D}) gives rise to a {\em compensated void},
\be\label{compensatedvoid}
\int_0^\infty dr\,r^2\,\delta(r,\tau) = 0\,,
\ee
a property which will be useful later. 

\begin{figure}
\centering
\includegraphics[width=0.9\columnwidth,angle=0]{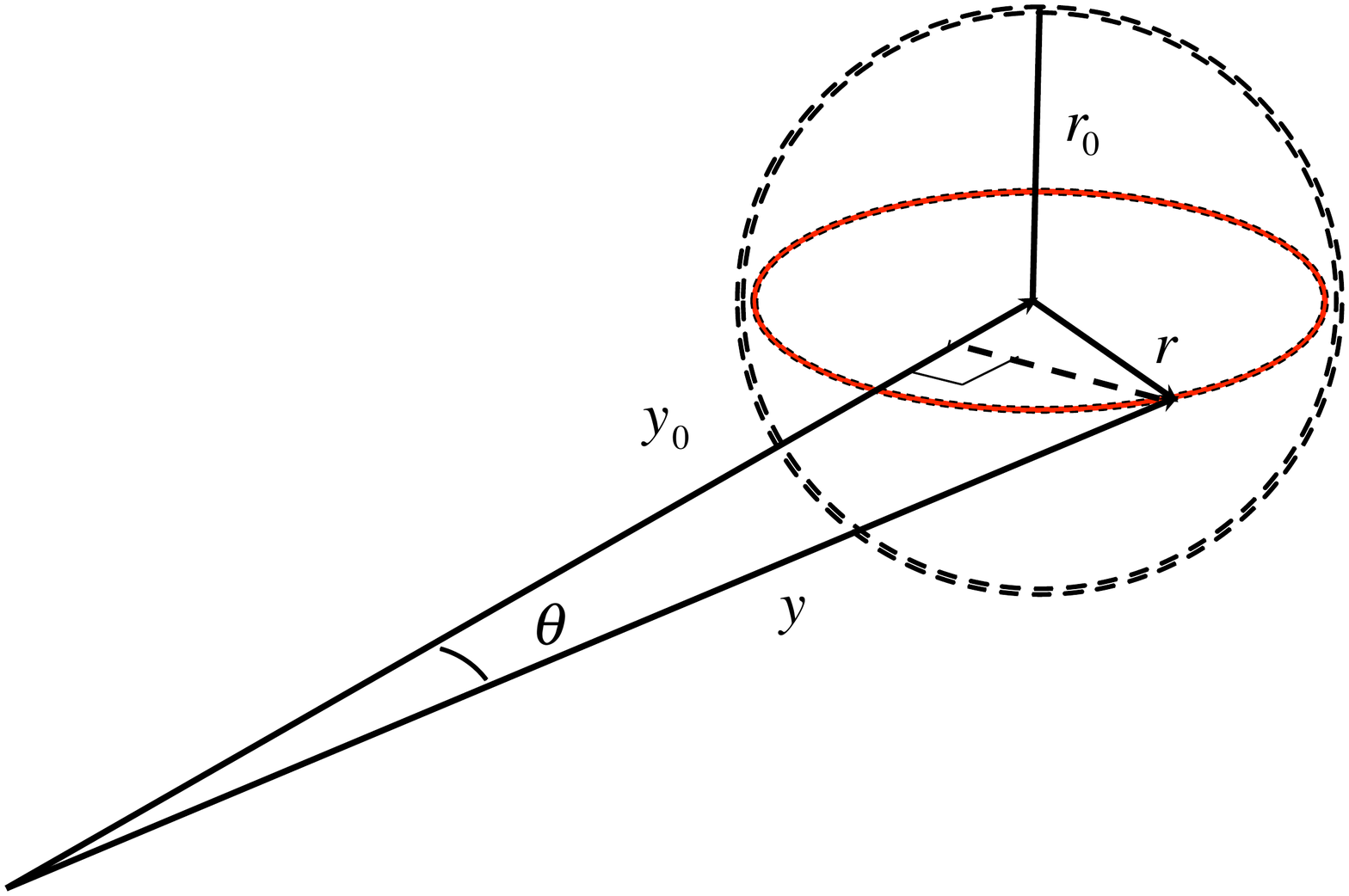} 
\caption{The corresponding geometry of the 2D projection of the LTB void for the
WISE-2MASS projected galaxy contrast. The observer is at a comoving distance $y_0$
from the center of the void.}
\label{2Dprojection}
\end{figure}

In order to compare with WISE-2MASS, we project the 3D density (\ref{profile3D}) onto the transverse plane, with the center of the void at comoving distance $y_0$, see Fig.~\ref{2Dprojection}. The
relation between $r$ and $\theta$ is given by $r^2(y,\theta)=y^2+y_0^2-2y y_0\,\cos\theta$,
\be
\delta_{2D}(\theta) = \int_0^\infty \delta \left(r(y,\theta)\right)\,y^2\phi(y)dy\,,
\label{profile_WISE}
\ee
with $y=r(z)$ and $y_0 = r(z_0)$ are the comoving distances to the void, and  $\phi(y)$ is the 
WISE-2MASS window function.

\begin{figure}
\includegraphics[width=8cm]{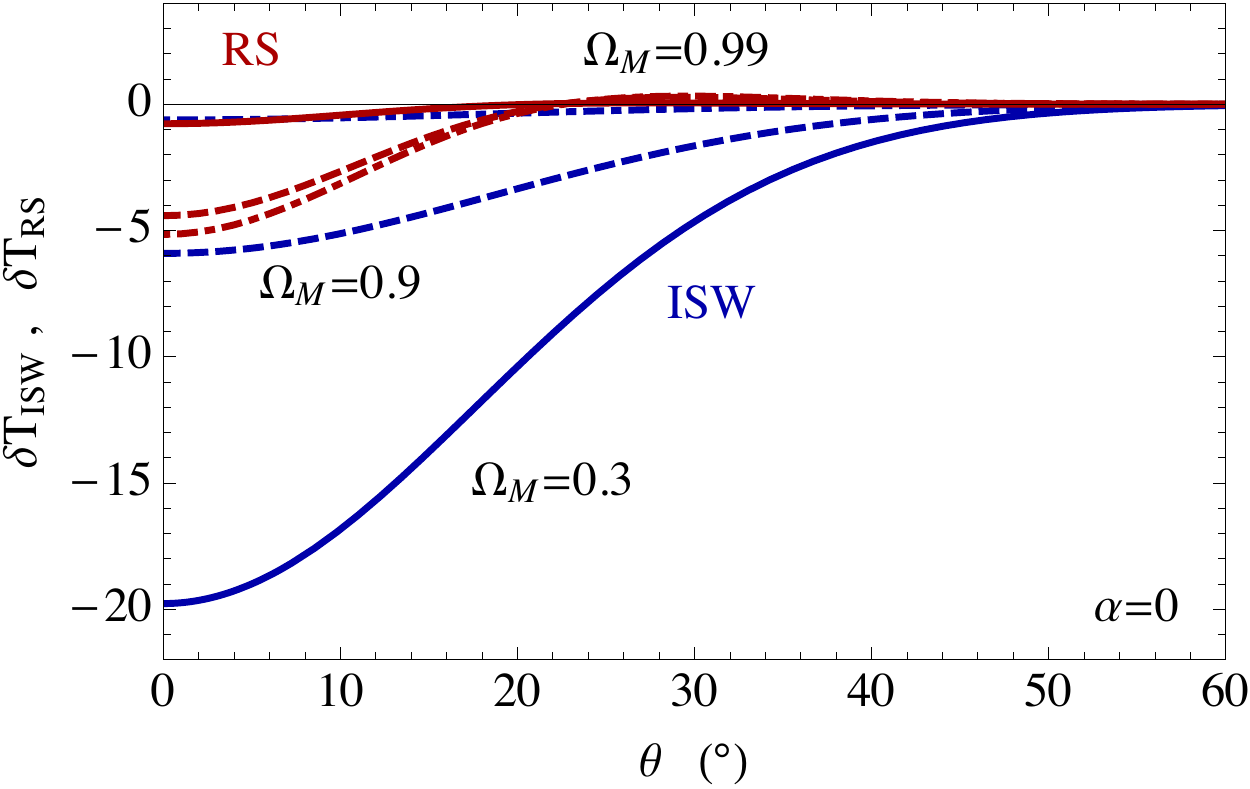}
\includegraphics[width=8cm]{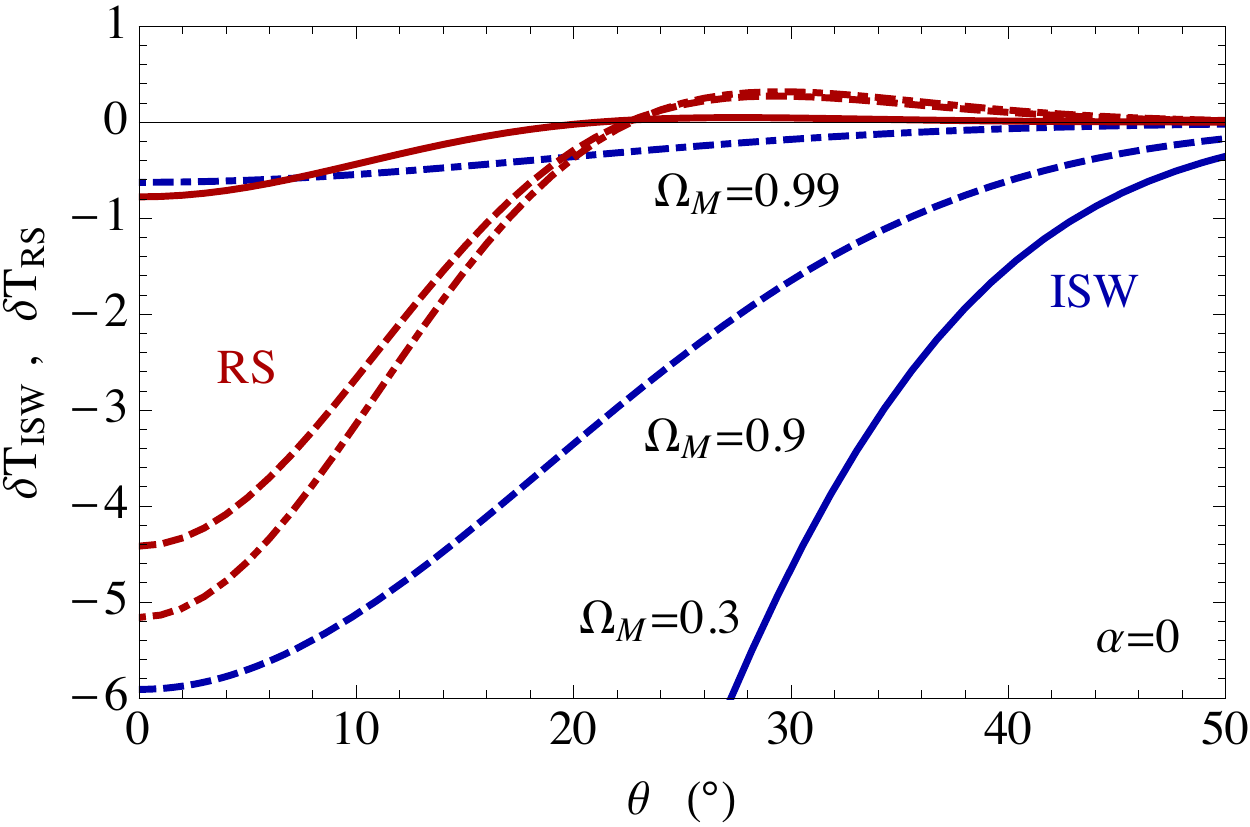}
\caption{In the top panel the ISW (blue) and RS (red) effect for LTB voids with the profile introduced 
in Eq. (\ref{profilekr_basic}) 
within our perturbative treatment are displayed 
for different values of $\Omega_M=0.3$ (solid), $\Omega_M=0.9$ (dashed), 
$\Omega_M=0.99$ (dot-dashed). In the bottom panel we zoom on the RS effect to show how  
its angular dependence differs from the one of the ISW term.
The parameters of the void are $r_0 = 195$ Mpc/h, $\delta_0=0.25$, $z_0=0.155$.}
\label{CMB_LTB_basic_pred}
\end{figure}

From the metric perturbation (\ref{profilePhi_basic}), we now compute the linear Integrated 
Sachs-Wolfe (ISW henceforth) \citep{ISW,ISW_lambda} and the non-linear 
Rees-Sciama (RS henceforth) \citep{RS} effect  on the CMB 
temperature anisotropies. For a large compensated void with a profile as in Eq.~(\ref{profile3D}), 
the linear ISW effect is given by
\ba
&&\hspace*{-2mm}\frac{\delta T}{T}^{\rm ISW} (\theta) = 2 \int_{\tau_0}^{\tau_E} d\tau\,\dot\Phi(\tau,r)  \\
&&\hspace*{2mm}= -\frac{12}{11}\frac{\OL}{\OM}\int_{0}^{1}\!da\,a^2\Phi_0(r)
\cdot {}_2\!F_1\Big[2,\,\frac{4}{3},\,\frac{17}{6},\frac{-\OL a^3}{\OM}\Big]  \nonumber \\
&&\hspace*{2mm}\equiv -\frac{12}{11}\frac{\OL}{\OM}\int_{0}^{1}\!da\,a^2\Phi_0(r(z,\theta)) \cdot
F_4(u)\,,  \nonumber
\ea
where $$r^2(z,\theta) = r^2(z) + r^2(z_0) - 2\, r(z)r(z_0)\cos\theta\,,$$ 
and we integrate over $z$. Note that the comoving distance is given by 
$r(z) = \tau_0 - \tau(z)$, see Eq.~(\ref{tau}).

For the Gaussian profile in Eq. (\ref{profilePhi_basic}) we find, in the small angle approximation, with $r(z_0) > r_0$,
\ba
&&\hspace*{-6mm}\frac{\delta T}{T}^{\rm ISW}\hspace*{-4mm}(\theta) \simeq -\frac{3\sqrt\pi}{22}
\frac{H(z_0)\,\OL\,F_4(-\OL/\OM(1+z_0)^3)}{H_0\,(1+z_0)^4\,F_1(-\OL/\OM)}\times  \nonumber  \\
&&\hspace*{-2mm}\Big(1+{\rm erf}\Big[\frac{z_0}{H(z_0)r_0}\Big]\Big)\,\delta_0\,(H_0r_0)^3
\exp\Big[\!-\frac{r^2(z_0)}{r^2_0}\theta^2\Big]\,, \label{dtisw}
\ea
which also gives a Gaussian profile for the ISW effect. 

Following \cite{Tomita,TomitaInoue}, we compute the RS effect  on the CMB 
temperature anisotropies as
\ba
\frac{\delta T}{T}^{\rm RS} (\theta) &\!=\!& \int_0^{z_{\rm LS}} dz
\left[4\zeta'_1(z)\frac{r^2(z,\theta)}{r^2_0} + 9\zeta'_2(z)\right]\times \\
&&\hspace*{15mm} \frac{100}{9} \frac{\Phi^2_0}{r_0^2} 
\exp\Big[\!-2\frac{r^2(z,\theta)}{r^2_0}\Big]\,,  \nonumber
\ea
where primes denote differentiation w.r.t. redshift and
\ba
&&\hspace*{-5mm}\zeta_1(z) =  \frac{3\,\tau^2}{200} \cdot \frac{F_1^2(u)}{F_2^2(u)}\,, \\
&&\hspace*{-5mm}\zeta_2(z) = \frac{-\tau^2}{210}\cdot \frac{1}{F_2^2(u)} \left[ 
\frac{7}{5}\left(\frac{5}{6}F_1(u)+\frac{1}{6}F_3^2(u)\right) \right. \\
&&\hspace*{23mm} \left. - \frac{2}{5}\sqrt{1+u}\,\left(F_1(u)F_3(u)-\frac{5}{12}G(u)\right) \right]\,,\nonumber\\
&&\hspace*{-5mm}G(u) = \frac{1}{3} u^{-7/6} \int \frac{\ du \, u^{1/6}}{\sqrt{1+u}}\,\left[F_3^2(u) - F_1(u)\right]\,.
\ea
Doing the integral in $z$, in the small angle approximation,
\ba
&&\hspace*{-5mm}\frac{\delta T}{T}^{\rm RS}\hspace*{-3mm}(\theta) \simeq - \frac{100}{9}\sqrt\frac{\pi}{8} 
\left(4\zeta'_1(z_0)\frac{r^2(z_0)}{r^2_0}\,\theta^2 + 9\zeta'_2(z_0) \right)H_0^2\times \nonumber \\
&&\hspace*{-2mm} \Big(1+{\rm erf}\Big[\frac{\sqrt2\,z_0}{H(z_0)r_0}\Big]\Big)\,
\frac{\Phi^2_0(r_0,\delta_0)}{H_0\,r_0} \exp\Big[\!-2\frac{r^2(z_0)}{r^2_0}\,\theta^2 \Big]\,.
\ea

The ISW and RS effects are shown in Fig.~\ref{CMB_LTB_basic_pred}.
The RS effect is quadratic in $\delta_0$ and cubic in $r_0$, 
but smaller than the ISW effect in a $\Lambda$CDM cosmology.\footnote{In a first version
of this manuscript we evaluate a much larger RS effect, mainly due to a wrong mismatch 
between the LTB and FRW radius, following \cite{MasinaNotari2009}. Our RS (and ISW) 
calculation for the novel profile in Eq. (\ref{profilekr_basic}) now agrees with \cite{Zibin,Nadathuretal}.}

\subsection{Predictions for the Draco Supervoid}

We construct a $\chi^2$ statistics corresponding to
the fits for the projected LTB model in the WISE-2MASS catalogue (denoted by the pedix LSS) 
and the corresponding effect on the {\sc Planck} temperature 
anisotropy pattern (denoted by the pedix CMB).  
As primary parameters, we chose 
$\{ \theta_i \}= \{\delta_0, r_0, z_0\}$, giving the following:
\ba
\chi^2_{_{\rm LSS}}\!&\!=\!&\!\sum_{ij} (\delta_{2D}(\theta_i) - \delta_i^{\rm LSS})D_{ij}^{-1}
(\delta_{2D}(\theta_j) - \delta_j^{\rm LSS})\,, \label{chi2lss} \\[1mm]
\chi^2_{_{\rm CMB}}\!&\!=\!&\!\sum_{ij} (\delta T(\theta_i) \! - \! \delta T_i^{\rm CMB})C_{ij}^{-1}
(\delta T(\theta_j) - \delta T_j^{\rm CMB})\,. \label{chi2cmb}
\ea
The term in Eq. (\ref{chi2lss}) corresponds to the $\chi^2$ of 
the projected LTB void profile (\ref{profile_WISE}) 
with respect to the observed WISE-2MASS galaxy distribution, 
using the highly correlated covariance matrix $D_{ij}$ of 
concentric rings in WISE-2MASS. We estimated the 
covariances by generating 10,000 Gaussian realizations of 
the projected WISE-2MASS map with {\tt Healpix synfast}. 
The simulations were created assuming {\sc Planck} cosmology, 
and the redshift distribution of the WISE-2MASS sources.

The term in Eq. (\ref{chi2cmb}) is the $\chi^2$ of the CMB profile compared to
the covariance matrix of rings in the CMB is also highly correlated. 
The covariance matrix was determined from 10,000 Gaussian CMB realizations. 
Note the covariance of the CMB at the angular scales considered is dominated 
by cosmic variance, thus neglecting the cross-correlation with LSS is a good approximation.

In order to match the data to the LTB prediction, 
we followed \cite{Kovacs2013} and \cite{SzapudiEtal2014}, and estimated 
the bias of the galaxy catalog using {\tt SpICE} \citep{spice} 
and {\tt python} CosmoPy\footnote{\texttt{http://www.ifa.hawaii.edu/cosmopy/}} 
package, finding $b=1.41\pm0.07$. 
The depression in galaxy counts, therefore, corresponds to a $\delta_{2D} = \delta_{2D,g} / b$ 
underdensity in matter, assuming linear bias relation.

According to Eq. (\ref{chi2lss}), the best-fit marginalized parameters to WISE-2MASS data alone 
is $\delta_0 = 0.40 \pm 0.20$, $r_0 = 210 \pm 70$ Mpc/$h$ at 68~\% CL. While
from Eq. (\ref{chi2cmb}), the best-fit to {\sc Planck} data alone is $\delta_0 = 0.23 \pm 0.40$, 
$r_0 = 290 \pm 90$ Mpc/$h$ at 68~\% CL \footnote{The $\chi^{2}$ statistic for the radial density profile measurement {\em compared
to a null value} in each bin, and found $\chi^{2}=65.22$ for 24 degrees-of-freedom, i.e. $p=10^{-5}$
as a simple estimate of the extremeness of the Draco supervoid in WISE-2MASS.}.

\begin{figure}
\centering
\includegraphics[width=8cm,angle=0]{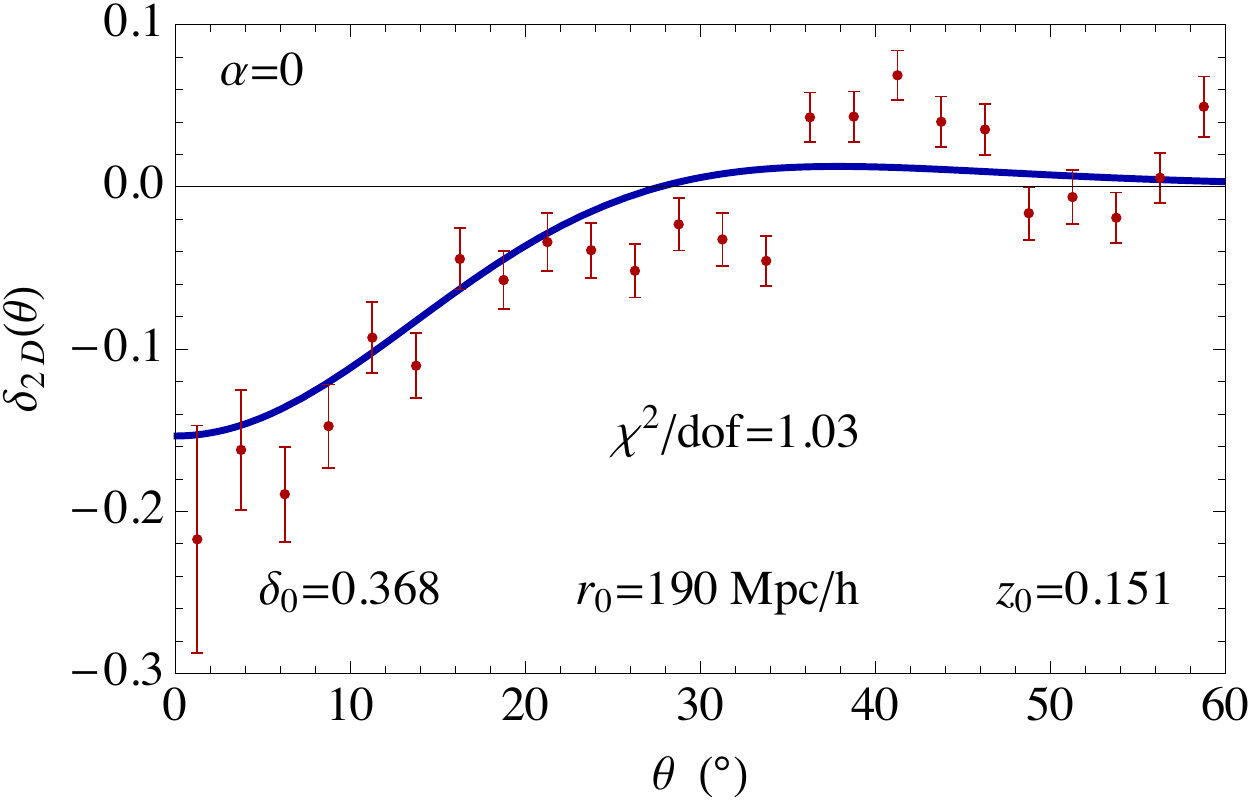} 
\vskip 10pt
\includegraphics[width=8cm,angle=0]{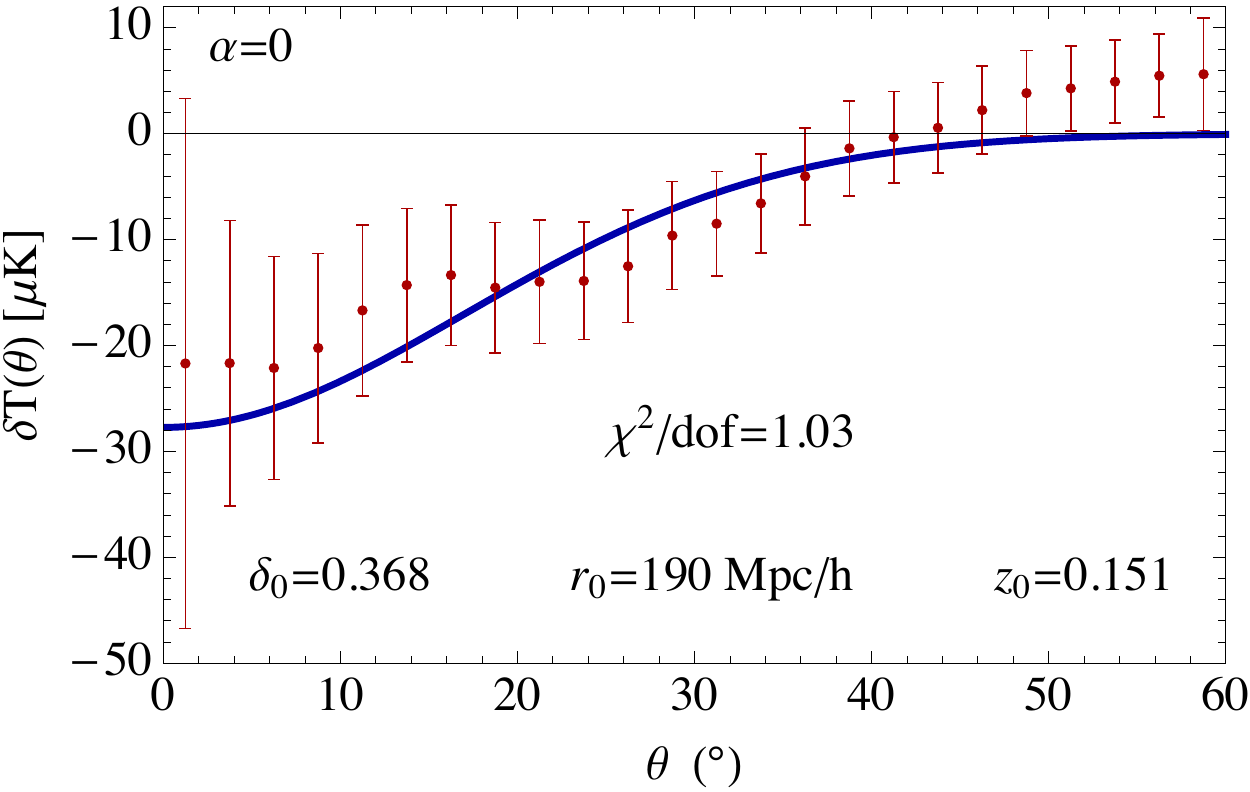}
\caption{The density profile from WISE-2MASS catalogue compared with the best-fit theoretical model
for the underdensity (\ref{profile_WISE}) from a combined analysis (top panel). The temperature profile
from {\sc Planck} SMICA map (bottom panel) is compared with the predicted CMB signal. 
The blue lines are the theoretical profiles for rings and in red are the measurements.
}
\label{v2_prof}
\end{figure}

\begin{figure}
\centering
\includegraphics[width=0.9\columnwidth,angle=0]{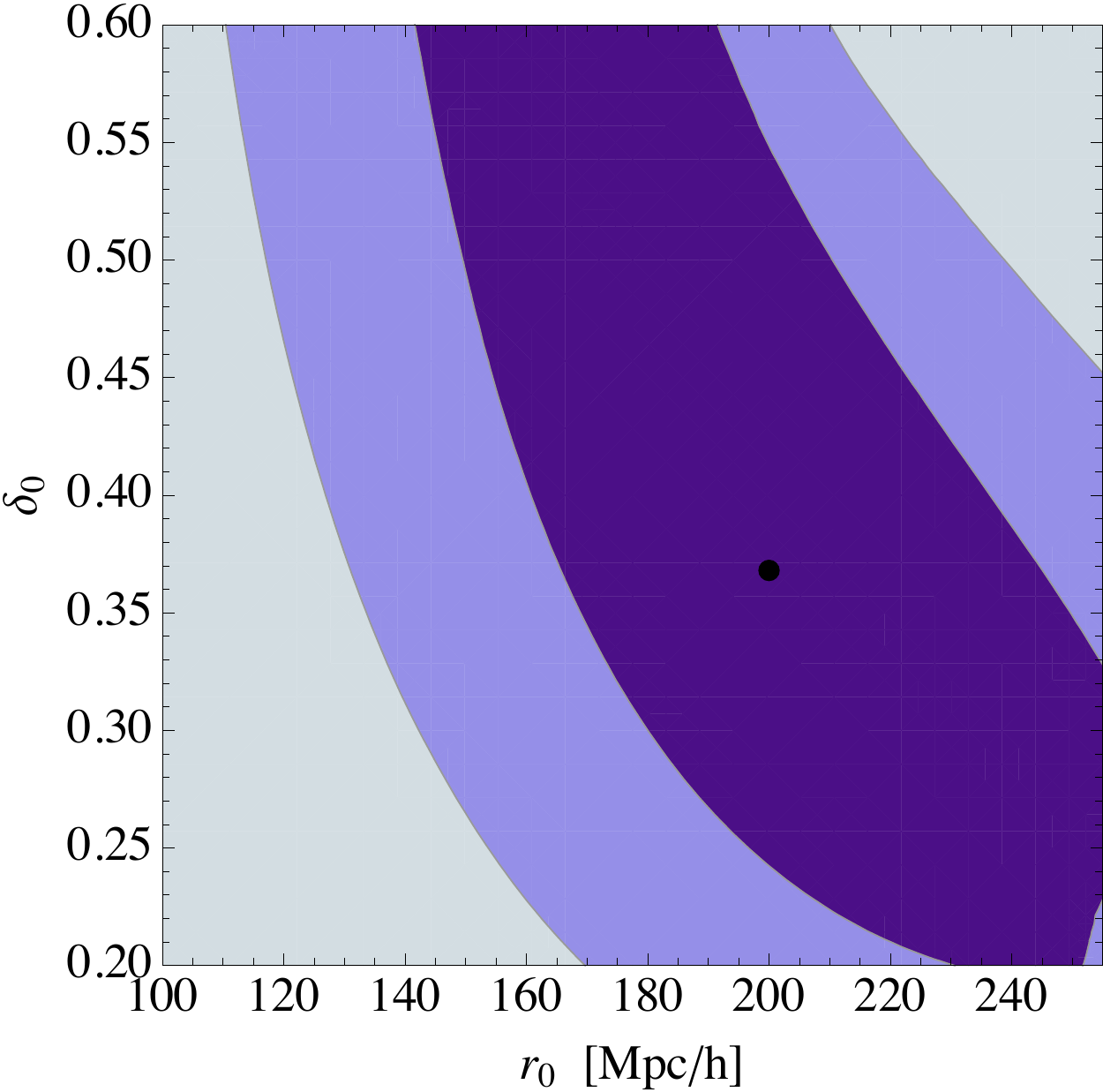}
\hspace{2mm}
\caption{The 1 and 2$\sigma$ contours of the size and depth parameters of the LTB void (\ref{profile3D}), 
after marginalization over the redshift of the center of the void with the WISE-2MASS window function.}
\label{contour_2CS}
\end{figure}

We now estimate a simultaneous fit of WISE-2MASS and {\sc Planck} data by:
\be
\chi^2_{_{\rm tot}}(\delta_0, r_0, z_0) = \chi^2_{_{\rm LSS}} + \chi^2_{_{\rm CMB}} \,.
\label{chi2tot} 
\ee
Simultaneous minimization yields the best fit parameters, which we quote here with marginalized 1$\sigma$ 
errors,
\ba
\delta_0\!&\!=\!&\! 0.37^{+0.22}_{-0.12} \ (1\sigma)\,,\\[2mm]
\hspace{1cm}r_0 ({\rm Mpc}/h)\!&\! =\!&\! 190^{+39}_{-27} \ (1\sigma)\,, \\[2mm]
z_0\!&\!=\!&\! 0.15^{+0.04}_{-0.05} \ (1\sigma)\,, 
\ea
which are in good agreement with the best-fits obtained separately from WISE-2MASS and Planck, 
and with the WISE-2MASS window function. The 1 and 2$\sigma$ contours of the size and depth parameters of the LTB void are shown in Fig.~\ref{contour_2CS}.

Note that the only constraint on the redshift of the void center relies on the matching with the CMB profile in the location of the supervoid, and assuming a particular LTB void model. The approximate central redshift of $0.10<z<0.20$ is just compatible with our prior knowledge that the void might be located relatively close to us \citep{Rassat2013}. This slight tension indicates that in reality the void might be smaller in physical size and its cooling effect on the CMB is less significant assuming the same LTB model.

For later comparison, we calculate the averaged underdensity within the best fit radius $r_0=190\,h^{-1}$Mpc. The 3D top-hat-averaged density from the LTB profile, see Eq.~(\ref{profile3D}), 
is $\bar\delta = 3/r_0^3  \int_0^{r_0}  r^2 dr \,\delta(r) = - \delta_0/e$. 
This finally gives the average void depth $\bar\delta = - 0.14 \pm 0.03$. 

In the $\Lambda$CDM model, the expected number of supervoids as extreme as the Draco supervoid at $z<0.5$ is roughly $N\sim10$ according to the estimations of \cite{Nadathuretal} in their Figure 5. However, for a more realistic modelling of the actual survey volume filled by the WISE-2MASS catalog, we need to take into account a masking factor $f_{sky}=0.53$, and the fact that the redshift distribution of the WISE-2MASS catalog prevents reliable void identification at $z\geq0.25$. The latter effectively results in a WISE-2MASS comoving survey volume that is only $\sim 15\%$ of the volume considered in \cite{Nadathuretal}. Including these factors, we estimate a probability of $p\approx0.8$ for finding a void at least as extreme as the Draco void in the WISE-2MASS survey volume.

\subsection{Modelling of the Cold Spot}

We now analyze a LTB description for the supervoid found in the CS direction. 
By using the results in section IV we obtain the marginalized over redshift best-fit 
to WISE-2MASS data with reduced $\chi^2 = 0.85$ and $\delta_0 = 0.29 \pm 0.19$, 
$r_0 ({\rm Mpc}/h) = 198 \pm 90$ at 68 \% CL \footnote{The $\chi^{2}$ statistic for the radial density profile
measurement {\em compared to the null value in each bin} is 
$43.94$ for 24 degrees-of-freedom, i.e. $p=0.007$.}. 
As best-fit to {\sc Planck} data we obtain a reduced 
$\chi^2 = 1.2$ and $r_0 ({\rm Mpc}/h) = 136 \pm 50$, at 68 \% CL, with no constraint on $\delta_0$.

As is clear from the discussion above, the angular profile of the temperature decrement in the CS 
cannot be properly  fit by the ISW+RS effect described in Section III. 
The CMB imprint computed in Section 3 does not exhibit any ridge about 15 degrees which instead characterizes the CS 
(\cite{ZhangHuterer2010} noticed this structure as an outer ring). 

\begin{figure}
\includegraphics[width=8cm]{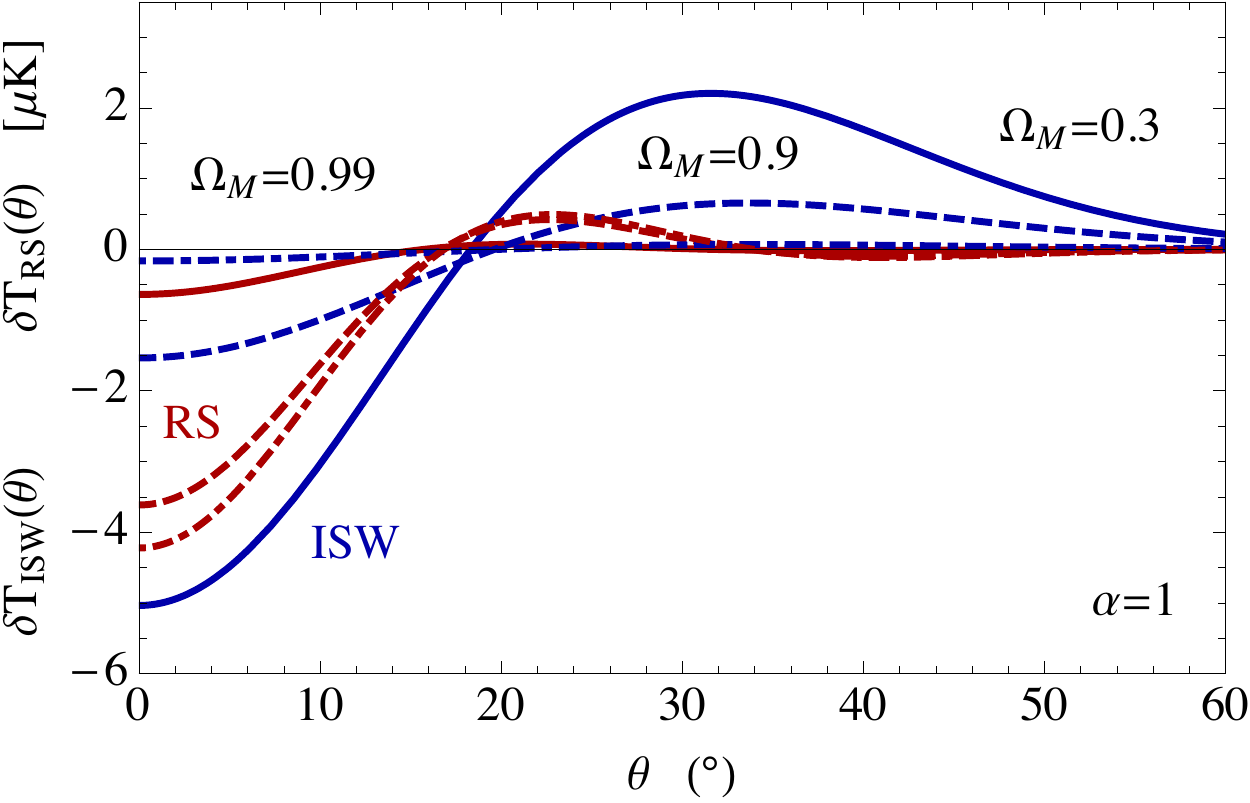}
\caption{In the top panel the ISW (blue) and RS (red) effect for LTB voids with the profile introduced in 
Eq. (\ref{profilePhi_modified})
within our perturbative treatment are displayed for different values of $\Omega_M=0.3$ (solid), $\Omega_M=0.9$ (dashed), $\Omega_M=0.99$ (dot-dashed). Note how for $\alpha=1$ the angular profiles of the ISW and RS effects still differ, but now the ISW term exhibits a profile similar to a compensated one. The parameters of the void are $r_0 = 195$ Mpc/h, $\delta_0=0.25$, $z_0=0.155$.}
\label{LTB_alpha_pred}
\end{figure}

We therefore modify the basic LTB profile by introducing a new parameter $\alpha$,
\be
\Phi_0(r) = \Phi_0\,\Big(1 - \al\,\frac{r^2}{r_0^2}\Big) \, \exp\Big[\!-\frac{r^2}{r_0^2}\Big]\,
\label{profilePhi_modified}
\ee
which corresponds to a density contrast:
\be
\delta(\tau,r) =  - \delta_0\,g(\tau) \Big( 1 - \frac{2+7\al}{3+3\al}\,\frac{r^2}{r_0^2} +
\frac{2\al}{3+3\al}\,\frac{r^4}{r_0^4} \Big) e^{-r^2/r_0^2}\,,
\label{profile3D_modified}
\ee
again giving rise to a {\em compensated void}~(\ref{compensatedvoid}), for all $\al$.
With these modifications, the ISW and RS angular profiles, for $0<\al<2$, are given by
\ba
&&\hspace*{-7mm}\frac{\delta T}{T}^{\rm ISW}\hspace*{-4mm}(\theta) \simeq -\frac{3\sqrt\pi}{22}
\frac{H(z_0)\,\OL\cdot F_4(-\OL/\OM(1+z_0)^3)}
{(1+\al)H_0(1+z_0)^4\cdot F_1(-\OL/\OM)}\times  \nonumber  \\
&&\hspace*{5mm} \left(\frac{2-\al}{2}-\al\,\frac{r^2(z_0)}{r^2_0}\theta^2\right)
\Big(1+{\rm erf}\Big[\frac{z_0}{H(z_0)r_0}\Big]\Big) \times   \nonumber \\
&&\hspace*{10mm} \delta_0\,(H_0r_0)^3
\exp\Big[\!-\frac{r^2(z_0)}{r^2_0}\theta^2\Big]\,, 
\ea
which now has a node and a positive maximum, and 
\ba
&&\hspace*{-5mm}\frac{\delta T}{T}^{\rm RS}\hspace*{-3mm}(\theta) \simeq -\frac{100}{9} \sqrt\frac{\pi}{8}
\left[9\zeta'_2(z_0)\Big(1+\frac{\alpha}{2} + \frac{7}{16}\alpha^2\Big) + \right. \nonumber \\
&& \hspace*{-5mm} 
+ \left(4\zeta'_1(z_0)\Big(1+\frac{\alpha}{2}\Big)
- 9\zeta'_2(z_0)\Big(2\alpha+\frac{\alpha^2}{2}\Big)\right)\frac{r^2(z_0)}{r^2_0}\,\theta^2 
+ \nonumber \\
&& \hspace*{5mm} 
+ \left.\Big(9\zeta'_2(z_0)\,\alpha^2 - 4\zeta'_1(z_0)\,\alpha\Big)\frac{r^4(z_0)}{r^4_0}\,\theta^4
\right] H_0^2 \times \nonumber \\
&&\hspace*{-5mm} \Big(1+{\rm erf}\Big[\frac{\sqrt2\,z_0}{H(z_0)r_0}\Big]\Big)\,
\frac{\Phi^2_0(r_0,\delta_0)}{H_0\,r_0} \exp\Big[\!-2\frac{r^2(z_0)}{r^2_0}\,\theta^2 \Big]\,.
\ea
These two effects are shown in Fig.~\ref{LTB_alpha_pred}. 
Note how this new compensated void model leads to a completely different 
angular profile for the ISW term, 
reminiscent of a compensated structure in the CMB as well, although with a smaller amplitude with respect to $\alpha=0$.

We choose from now on the parameter $\alpha=1$. 
We then obtain a best-fit to WISE-2MASS data with reduced $\chi^2 = 0.85$ (for 24 d.o.f.) 
and $\delta_0 = 0.27 \pm 0.25$, $r_0 ({\rm Mpc}/h) = 270 \pm 90$ at 68 \% CL. 
By fitting to {\sc Planck} data alone we obtain a similar radius, $r_0 ({\rm Mpc}/h) = 254 \pm 50$ at 68 \% CL, 
with a large underdensity, $\delta_0 > 0.6$ at 95 \% CL, which is clearly too high for the $\Lambda$CDM model.

Our $\Lambda$LTB models for the WISE-2MASS underdensity and the CMB decrement in 
the direction of the CS are not as satisfactory as for the analogous alignment 
we have found in the direction of the Draco Supervoid. 
Although we have introduced a second novel compensated LTB profile, which leads 
to an ISW plus RS angular profile reminiscent of the CMB decrement 
in the direction of the CS, we cannot find an amplitude which fits well WISE-2MASS and 
{\sc Planck} simultaneously for $\alpha=1$. 

Given the current status of the WISE-2MASS observations, 
we cannot exclude that other values of $\alpha$ or other profiles could provide a better fit.
Another issues are the approximations used in this paper, as the perturbative treatment 
in a flat FRW metric, the assumption of $\Lambda$CDM, or spherical symmetry.
It is conceivable that the structure in the direction of the CS might not 
be well approximated by a spherical object and might be possibly elongated 
along the line of sight: in such a case the angular profile in the WISE-2MASS 
data would be smaller than the effective distance which enters in the CMB decrement. 
Although this calculation goes beyond the scope of the current paper, this different 
geometry for the underdensity would help in reconciling the amplitude of the CMB decrement with the underdensity seen in WISE-2MASS. 

\begin{figure}
\centering
\includegraphics[width=8cm,height=5cm,angle=0]{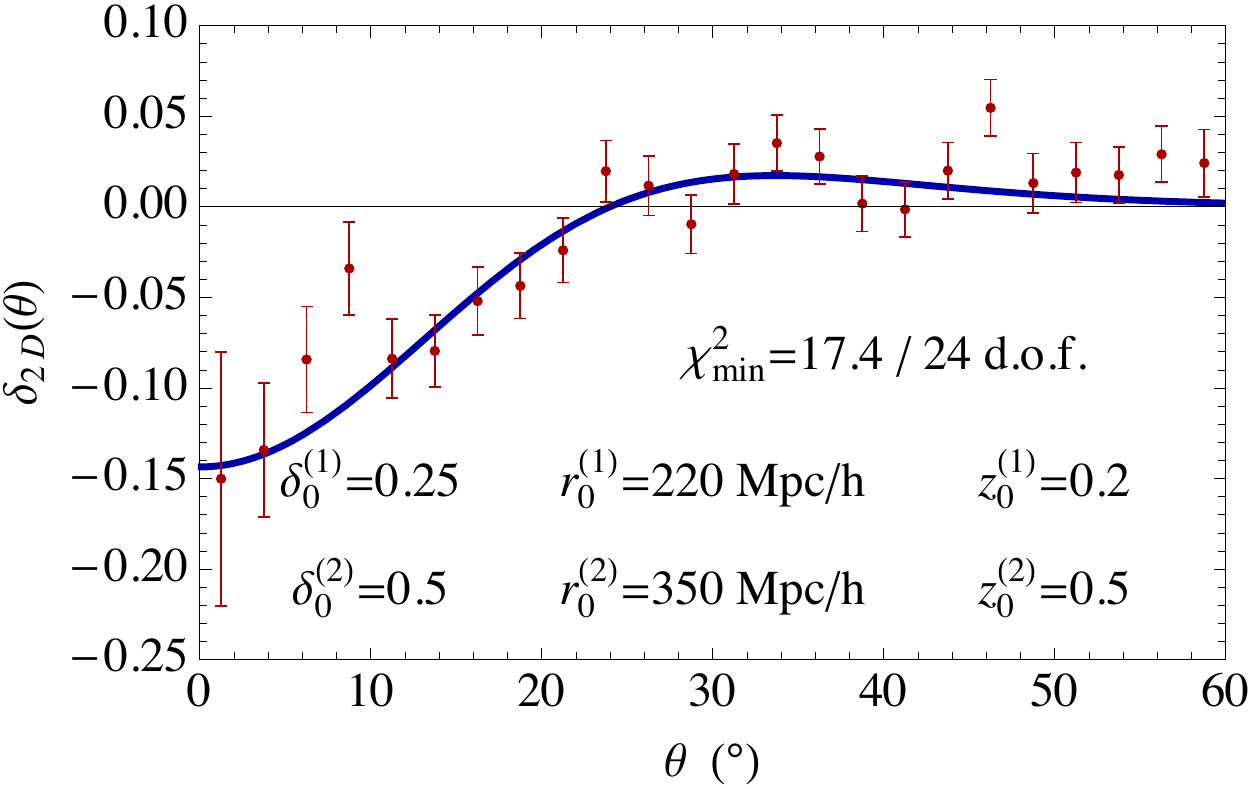}
\vskip 10pt
\includegraphics[width=8cm,height=5cm,angle=0]{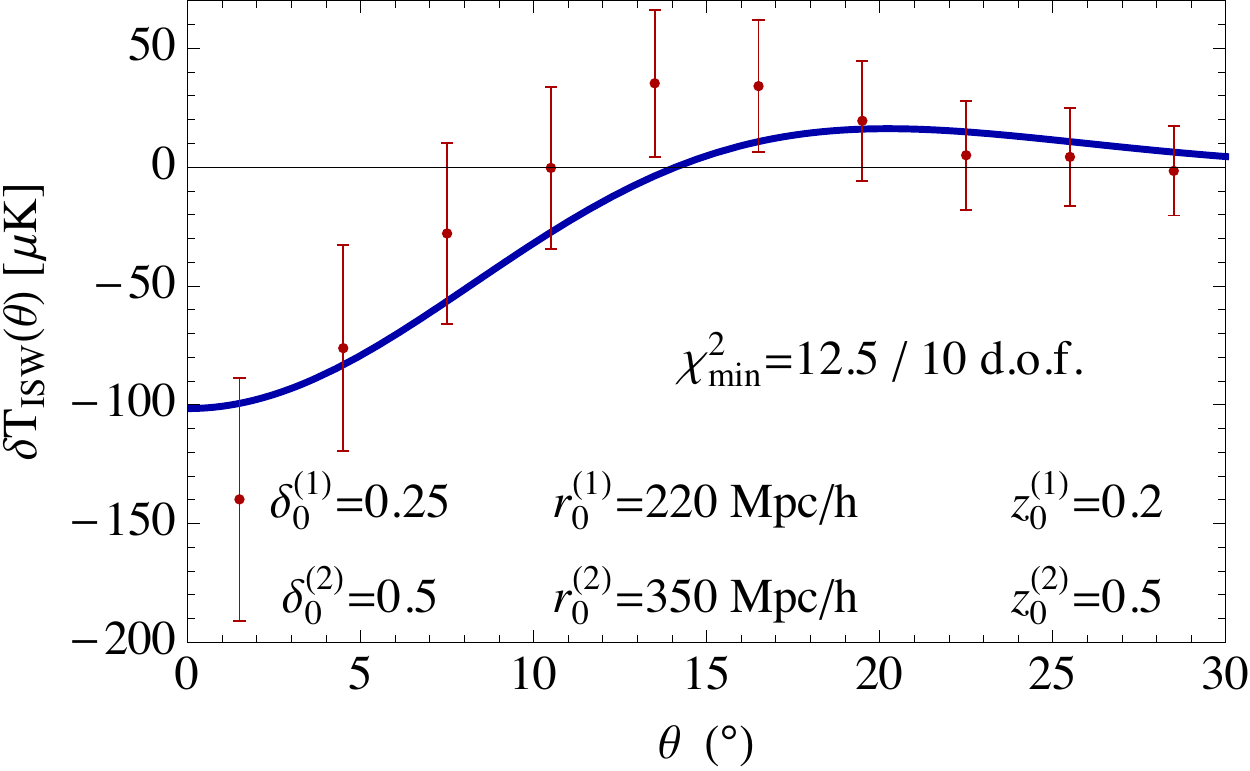}
\caption{The density profile from WISE-2MASS catalogue compared with the theoretical model
for the underdensity (\ref{profile_WISE}) (top panel). The temperature profile from {\sc Planck} SMICA map (bottom panel) is compared with the predicted CMB signal. The blue lines are the theoretical profiles for rings and in red are the measurements. The model considers two adjacent spherical voids along the same line of sight with different sizes and depths.
}
\label{2void_profiles}
\end{figure}

In fact, if we consider together two adjacent voids 
along the same line of sight, at two different redshifts, 
it is possible to find a reasonable minimum $\chi^2$, 
at the expense of an unrealistic depth (i.e. high density contrast) 
for the furthest void, although with reasonable widths in both, 
see Fig.~\ref{2void_profiles}. The common reduced $\chi^2=0.96$ 
for the first void at $z_0=0.2$ with a width $r_0 = 220$ Mpc/$h$ 
and depth $\delta_0 = 0.25$, and for the 
second void at  $z_0=0.5$ with a width $r_0 = 350$ Mpc/$h$ and depth $\delta_0 = 0.5$ with reduced $\chi^2=17.4/24$ d.o.f. for WISE-2MASS and
$\chi^2=12.5/10$ d.o.f. for {\sc Planck}.

For estimations of the cosmic rareness of the CS void, 
we rely on the (presumably more accurate) best-fit void parameters 
obtained with photo-z mapping by the companion paper \cite{SzapudiEtal2014} 
($\delta_{2D} = 0.14$, $\delta_0 = 0.38$, $r_0 = 220$ Mpc/$h$). 
The probability of detecting supervoids at least as extreme as the 
CS supervoid at $z<0.5$ is roughly $p\sim0.1$ according to the 
estimations of \cite{Nadathuretal}. Including the $f_{sky}=0.53$ 
and $V_{z<0.25}/V_{z<0.5}\approx0.15$ factors, the estimated probability is $p\sim0.01$ 
for finding a void at least as extreme as the CS void in the WISE-2MASS survey volume. 
Thus the CS void represents a rare fluctuation in $\Lambda$CDM, and its occurrence is 
less likely than that of the Draco Supervoid. 
Note that the large errors on $\delta_{0}$ and $r_{0}$ 
can result in drastic changes in the expected void 
probability and ISW-RS signal of such voids \citep{Nadathuretal}. We, therefore, refer to these probabilities as crude estimations of cosmic rareness.

\section{Discussion \& Conclusions}

We have found two large underdensities in the WISE-2MASS catalogue \citep{KovacsandSzapudi2013} smoothed on $20^\circ$ scales. 
The Draco supervoid is located at {$(l,b)\approx(101^\circ,46^\circ)$} and it corresponds to a gentle decrement in the CMB 
observed by both WMAP and {\sc Planck}, and it is well explained by the ISW and RS effects. The other 
supervoid is in the well known direction of the CMB CS, and its explanation in terms of ISW and RS effect is less satisfactory. 
Nevertheless, observationally, we found that the two most prominent supervoids identified in a full sky catalog coincide with cold areas of the CMB.

A related project \cite{SzapudiEtal2014} used the same WISE-2MASS data set supplemented with
PS1 photometric redshifts and the data by \cite{Granettetal2010} 
for a direct tomographic imaging of the CS region. They found a 
supervoid at $z = 0.22\pm 0.03$ with radius $r_0 = 220\pm 50\,h^{-1}$Mpc and depth of $\delta = -0.14 \pm 0.04$.

Motivated by these findings, we developed novel spherically symmetric $\Lambda$LTB
compensated void models to explain the CMB decrements 
observed in the direction of the two largest underdensities 
(supervoids) of the WISE-2MASS catalog. We compute perturbatively 
the projected angular density profile, and the ISW and RS angular 
profiles following the second order treatment presented in \cite{Tomita,TomitaInoue}.  

Within our perturbative treatment of the simplest family of LTB voids introduced here 
with a Gaussian profile for the curvature, we find that the Integrated Sachs-Wolfe and Riess-Sciama effects due 
to the Draco Supervoid found in the WISE-2MASS catalogue can explain well the CMB decrement 
observed in the same direction. By considering the combined fit with WISE-2MASS catalogue and {\sc Planck} we 
obtain $\delta_0 = 0.37^{+0.22}_{-0.12}$, $r_0 ({\rm Mpc}/h) = 190^{+39}_{-27}$, $z_0 = 0.15^{+0.04}_{-0.05}$ 
at 1$\sigma$ as a best-fit. The estimated redshift of the underdensity is in good agreement with the WISE-2MASS 
window function, but 2MASS-only observations of this underdense region by \cite{Rassat2013} indicate that the 
Draco void might be located closer to us, and therefore slightly smaller in physical size.

The explanation of the large decrement for the CS in terms of the ISW and RS effects is known to be difficult \citep{Inoue2006,Inoue2007}. The CMB angular profile of the CS, studied here at larger angular distances than those 
commonly considered ($\sim 10^\circ$), constitute also a puzzle with respect to the standard angular profiles previously 
studied \citep{Inoue2006,Inoue2007} and those obtained here with the perturbative treatment of the simplest class of LTB voids.
By introducing a further generalization of the LTB voids, which allows a more radical compensation, 
we indeed find an angular profile which better mimics what is observed in the CS, but with a much smaller amplitude for the 
underdensity found in the redshift range of the WISE-2MASS catalogue.

Given the status of the LSS observations and the assumptions used in this paper, 
more precise observations and further theoretical developments are needed before the CMB CS will be satisfactorily explained. 
On the observational ground, the 21,200 ${\rm deg}^{2}$ photometric redshift catalog of WISE-2MASS 
galaxies matched with SuperCOSMOS data (Kov\'acs et al., in preparation) 
will provide an enhanced tool for examining these structures and for identifying more superstructures in the low-$z$ Universe.

Concerning the assumption of $\Lambda$CDM, there is some debate about the effects on the CMB being stronger than the simple predictions of the linear ISW model in a $\Lambda$CDM cosmology \citep[e.g.,][]{GranettEtal2008,GranettEtal2009,PapaiSzapudi2010,CaiEtal2013}.
We also mention how the assumption of spherical symmetry or the use of perturbation theory or the absence of cross-correlation between the SW and ISW terms in our calculation could be important quantitative aspects, which go beyond the scope of the present project.

Despite the assumptions and caveats mentioned above, let us finally note that this work, and the related project \cite{SzapudiEtal2014}, significantly increased the probability of a physical explanation of the CS instead of a statistical fluke.
Given the rareness of the supervoid found in the CS direction, estimated here at about a 
3$\sigma$ deviation, we could ask for the probability of a random alignment of such a supervoid in the WISE-2MASS with a rare fluctuation in the CMB as the Cold Spot. 
Even considering the CS only as a 2$\sigma$ fluctuation, 
the hypothesis of this alignment as a statistical fluke is quite low. We
presented a quasi-linear ISW term with a angular profile similar to the CS and an amplitude 
larger than 20 $\mu$K  obtained within spherical symmetry. Thus a fairly typical cold primordial fluctuation enhancing the ISW effect from the CS void on the CMB sky could provide a plausible explanation.

\section*{Acknowledgements}

FF and JGB wish to thank the bilateral 
agreement INFN-CICYT and in particular the 
director of the Bologna INFN section 
Dr. G. Bruni for partially supporting this project. 
FF acknowledges the support by 
ASI through ASI/INAF Agreement I/072/09/0 for the {\sc Planck}
LFI Activity of Phase E2 and 
by MIUR through PRIN 2009 grant no. 2009XZ54H2. 
JGB acknowledges financial support from the Madrid 
Regional Government (CAM) under the program HEPHACOS S2009/ESP-1473-02, 
from the Spanish MICINN grants AYA2009-13936-C06-06, 
FPA2012-39684-C03-02 and Consolider-Ingenio 2010 PAU (CSD2007-00060), 
and from the MINECO Centro de Excelencia Severo Ochoa Programme, 
grant SEV-2012-0249. In addition, 
AK takes immense pleasure in thanking the support provided by the Campus Hungary fellowship program. Funding for this project was partially provided by the Spanish Ministerio de Econom'a y Competitividad (MINECO) under projects FPA2012-39684, and Centro de Excelencia Severo Ochoa SEV-2012-0234. 
IS acknowledges support from NASA grants NNX12AF83G and NNX10AD53G.

\end{document}